\documentclass[a4paper, 10pt, oneside,number,sort&compress, 3p]{elsarticle}

\journal{Nuclear Instruments and Methods A}

\usepackage{graphicx}
\usepackage{dcolumn}
\usepackage{bm}

\usepackage[english]{babel}
\usepackage[latin1]{inputenc} 

\usepackage{epstopdf}
\usepackage{amsfonts}
\usepackage{color}
\usepackage{epsfig}
\usepackage{mathrsfs}
\usepackage{amsmath,amssymb}
\usepackage{subfigure}

\usepackage{framed}
\usepackage{lipsum}
\usepackage[svgnames]{xcolor}
\definecolor{shadecolor}{named}{LightGrey}

\usepackage{float}
\floatstyle{plaintop}
\restylefloat{table}
\usepackage[tableposition=top]{caption}

\usepackage{blindtext}

\usepackage{fancyhdr}
\usepackage{multicol}

\usepackage{pifont}%
\usepackage{geometry}%
\usepackage{fleqn}
\usepackage{txfonts}
\usepackage{hyperref}

\def\bx{\mathbf{x}}

\def\bz{\mathbf{z}}
\def\bU{\mathbf{U}}
\def\bX{\mathbf{X}}

\def\bZ{\mathbf{Z}}
\def\bdelta{\mbox{\boldmath $\delta$}}
\def\bdeltaT{\mbox{\boldmath $\delta^{T}$}}

\def\bmu{\mbox{\boldmath $\mu$}}
\def\bxi{\mbox{\boldmath $\xi$}}

\def\bTheta{\mbox{\boldmath $\Theta$}}
\def\btheta{\mbox{\boldmath $\theta$}}
\def\bSigma{\mbox{\boldmath $\Sigma$}}

\def\bOmega{\mbox{\boldmath $\Omega$}}

\def\bZero{\mbox{\boldmath $0$}}

\def\cL{\mathcal{L}}

\usepackage{lineno}

\def\Xmax{\ifmmode {X_\mathrm{max}}\else
                   {$X_\mathrm{max}$}\fi\xspace}%

\begin{document}

\begin{frontmatter}

\title{Handling missing data in a neural network approach\\for the identification of charged particles in a multilayer detector}

\author[INAF]{S. Riggi\corref{cor}}
\ead{sriggi@oact.inaf.it}\cortext[cor]{Corresponding author.}
\author[Keras]{D. Riggi}
\author[FisicaUniCT]{F. Riggi}

\address[INAF]{INAF - Osservatorio Astrofisico di Catania, Italy}%
\address[Keras]{Keras Strategy - Milano, Italy}
\address[FisicaUniCT]{Dipartimento di Fisica e Astronomia - Universit\`{a} di Catania, and INFN, Sezione di Catania, Italy}

\begin{abstract}
Identification of charged particles  in a multilayer detector by the energy loss technique may also be
achieved by the use of a neural network. The performance of the network becomes worse when a large fraction of
information is missing, for instance due to detector inefficiencies.
Algorithms which provide a way to impute missing information have been developed over the past years. 
Among the various approaches, we focused on normal mixtures models in comparison with standard mean imputation and multiple imputation methods.
Further, to account for the intrinsic asymmetry of the energy loss data, we considered skew-normal mixture models and provided 
a closed form implementation in the Expectation-Maximization (EM) algorithm framework to handle missing patterns.
The method has been applied to a test case where the energy losses of pions, kaons and protons in a six-layers Silicon detector are
considered as input neurons to a neural network. Results are given in terms of reconstruction efficiency and purity of
the various species in different momentum bins.
\end{abstract}


\end{frontmatter}

\section{Introduction}\label{IntroductionSection}
\begin{figure}[!t]
\centering
\includegraphics[scale=0.4]{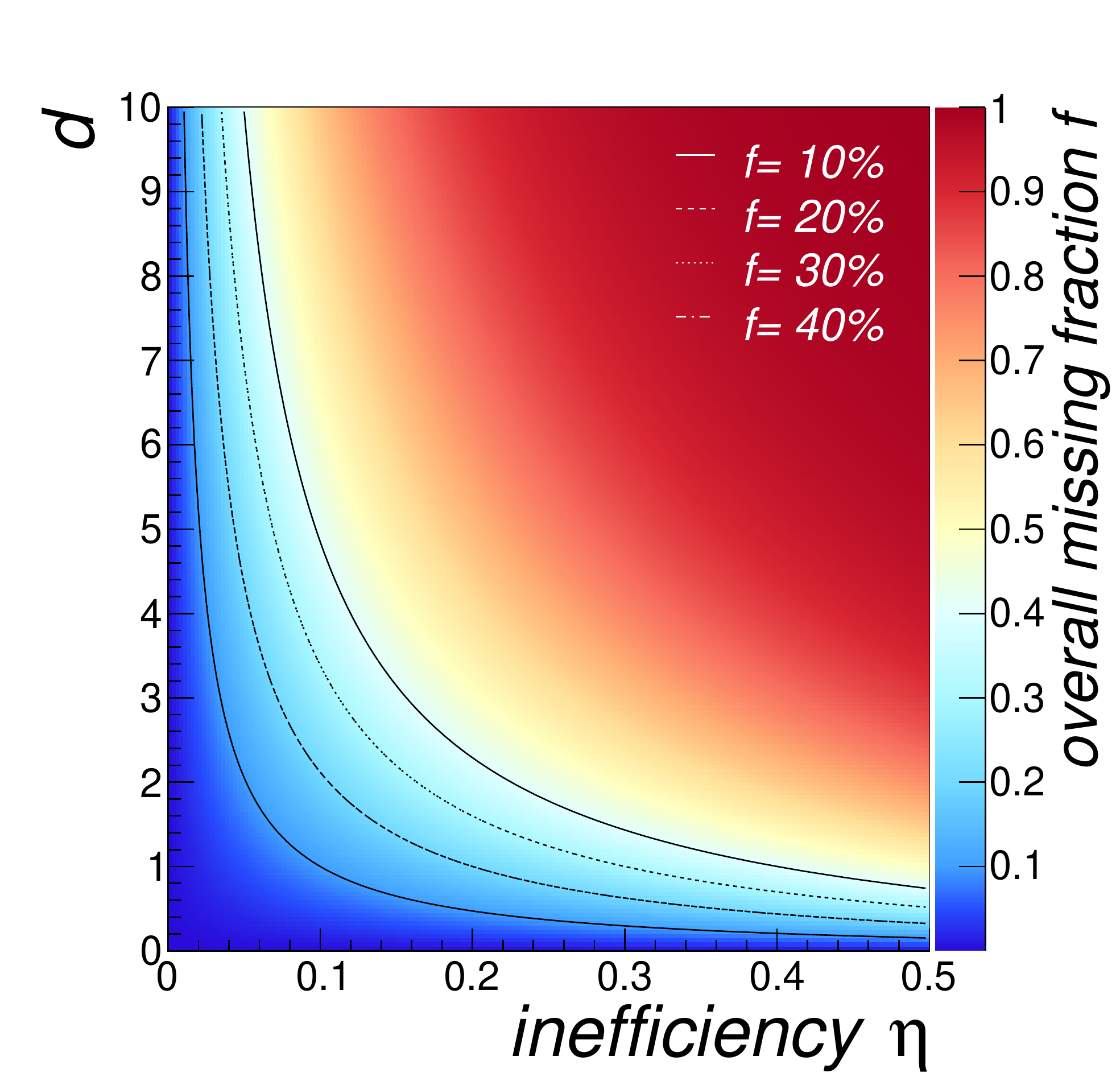}%
\vspace{0.1cm}
\caption{Overall missing fraction $f$ as a function of the variable detection inefficiency $\eta$ and the number of 
observables $d$ in the analysis. The contour lines are relative to $f$=10\%, 20\%, 30\%, 40\%.}
\label{OverallMissingRateVSIneffVSNDim}
\end{figure}

The treatment of missing data in different areas of science, statistics, economics, or pattern recognition has been and still is a wide subject 
of interest, since a variety of situations may lead to incomplete data in a set of information.
 In particle and nuclear physics there are  several situations where a certain fraction of data may be missing. 
Typical cases are the set of space points which contribute to the tracking of charged particles, or the different values of 
the energy loss of particles in a multilayer detector. 
There are several reasons why data may be missing.
In the simplest case, they may be missing completely at random (MCAR), e.g. the probability that a data is missing does not depend on the value 
of the variable. In most cases however such probability either depends on the other variables in the data set or on the value of the variable under consideration. 
In the former situation the data are said to be missing at random (MAR), while in the latter situation the missing data mechanism is denoted as non-ignorable or 
missing not at random (MNAR).
Examples of the MCAR case are the passage of minimum ionizing charged particles
through a multilayer detector, in the limit of very small thickness (thus limiting the amount of energy loss and multiple scattering in each layer),
with detection efficiency smaller than 100\%, due to dead areas or other inefficiencies. 
Examples of the MNAR case are given by a detector which has a finite energy threshold, modelled for instance as a sigmoid function.\\
The simplest approach in case of a certain fraction of missing information from a set of variables is to disregard the event where at least one variable is 
missing. Such approach, although retaining only complete events, may lead to a substantial loss of events either when the elementary fraction of missing 
events is large or when the number of variables is large. As an example, in a process with $d$=100 variables (such as the number of 
space points in a large tracking detector), 
even an elementary fraction $\eta$= 0.1\% in each variable leads to a net loss of 1-(1-$\eta$)$^{d}$ = 1-(0.999)$^{100}\sim$10\%.
Figure \ref{OverallMissingRateVSIneffVSNDim} shows the contour lines corresponding to different overall fractions of missing events 
(10\%, 20\%, 30\% and 40\%), as a function of the dimensionality $d$ of the problem and of the elementary fraction of missing events $\eta$ in each variable.
The importance of using all collected events is of special concern in case of rare events, where disregarding the event due to its incompleteness may
lead to a substantial fraction of potentially interesting events being lost. This is the reason why different methods have been employed
to impute the missing values of a variable according to the statistical properties of the variables of interest which define the
event \cite{Graham2009,Horton2007,Luengo2010}. Most of such methods rely on normal distributions of the variables, which however is not a  good
representation in many problems. The energy loss of a charged particle in a thin detector is a typical example of a variable whose distribution
has an asymmetric shape (Landau tail). For such reason, it is of interest to develop and test methods which are not biased by the assumption
on normal distributions. \\
In this paper an approach based on multivariate skew-normal (MSN) distribution was developed. The method was then applied to a test case, where
simulated energy losses of pions, kaons and protons in a multilayer Silicon detector were considered. The performance of a neural network was then
evaluated under missing data and after recovering them with the approach described here. Sect. \ref{MissingDataSection} describes in some more detail the problem of
missing data reconstruction and the methods adopted throughout the paper, namely Multiple Imputation (MI) and Maximum Likelihood (ML) methods, 
while the application to the test case is discussed in Sect. \ref{MethodApplicationSection}. Various methods
of imputation of the missing data were then considered and applied to the same set of simulated data, comparing their performance in terms of
identification efficiency. The results of such comparison are reported in Sect. \ref{ResultsSection}. Some detail of the algorithms being employed are described 
in \ref{SkewNormalAlgorithmDerivation} .

\section{Missing Data Reconstruction}\label{MissingDataSection}
The most common strategy of dealing with missing data is list-wise deletion (LD), that consists in rejecting all events with missing observables.
Such approach decreases the effective sample size, specially with a large number of observables in analysis, and correspondingly the 
power of any statistical tests to be performed with that sample, which in presence of rare events it is not desirable. 
If the data are not missing at random, such procedure may also lead to a selection bias. In cases when the missing data mechanism 
is MCAR or MAR list-wise deletion does not add any bias and if the event sample size is not a critical issue it should be the preferred approach.\\
The alternative approach than removing events with missing data, is to fill in or impute the missing values, 
with the aim of recovering the relevant features of the data set as a whole (mean, variances, parameters and so forth), rather than 
obtaining precise estimates of single missing values. Indeed, imputed values should not be trusted nor directly used to draw inferences.\\
The simplest way is to replace each missing value with the mean of the observed values for that variable. This strategy significantly alters the 
distribution for that observable and all derived summary indicators, for example the variance which is typically underestimated. 
If more than one group is present in the data sample, the mean estimation tend to be pulled towards the most abundant group. 
For that reason such method is not recommended.\\
A wide number of modern imputation methods exists in literature also used in the context of neural network analysis, 
among them regression-based single imputation, multiple imputation (MI), maximum likelihood (ML) methods, methods based on K-Nearest Neighbor (KNN) 
or K-means clustering, Singular Value Decomposition (SVD), Support Vector Machines (SVM). A review of some of these methods is available in 
\cite{Graham2009,Horton2007,Luengo2010}.\\Neural networks themselves could be also used to impute missing data. 
This is typically achieved by designing a set of neural network classifiers, with each specialized network performing classification on a different 
set of input attributes. However when the dimension of the input vector is large and missing data are present in many variables, such approach lead to a 
large number of combinations of networks to be implemented.\\In this work we will adopt maximum likelihood methods based on multivariate normal (MN) mixtures 
as compared to the popular multiple imputation approach.
Both methods assume a normal model, which is not appropriate to the energy deposit data under consideration. For that reason we designed in this work 
a ML method based on multivariate skew-normal (MSN) mixture models, presented in paragraph \ref{SkewNormalImputationSection}.

\subsection{Multiple Imputation}\label{MultipleImputationSection}
Single imputation methods, for example those based on linear regression, are intrinsically limited.
They infact proceed by calculating the regression of the incomplete variable on the other complete variables and
substituting the predicted mean for each missing observation. Since imputed values always lie on the regression line, the actual dispersion of the data is 
ignored and therefore the variance is underestimated, leading to a bias in the parameter estimates.\\The multiple imputation approaches proceed instead 
by introducing a random variation in the process, i.e. a random normal error into the regression equation, and by generating several data sets, each 
with different imputed values, to partially restore the lost variance. To account for the fact that only a single draw from the data population is taken, 
multiple random draws from the posterior distribution of the population, each imputed multiple times, are also introduced to completely restore the 
variance of the data. Multiple imputation algorithms available in literature typically differs for how this latest step is performed. 
Some uses bootstrap procedures to generate random draws, others adopt Data Augmentation (DA) techniques or the Markov chain Monte Carlo (MCMC) algorithm and so forth.
If the MAR assumption holds the method leads to almost unbiased estimators.

\subsection{Maximum Likelihood Imputation}\label{LikelihoodImputationSection}
In the likelihood approach a model is assumed to describe the observed data. If $p$ variables are considered for the analysis, we denote a sample of $N$ data
observations with $\bx$= ($\bx_{1}$, $\bx_{2}$, \dots, $\bx_{N}$), being $\bx_i$ a $p$-dimensional data vector. 
For a given observation $i$ we may have missing patterns. We indicate with $\bx_{i,o}$ the observed patterns and with $\bx_{i,m}$ the missing patterns.
A widely used approach consists in assuming a mixture model F($\bx$;$\bTheta$) with $K$ components with probability density functions \emph{pdf} f($\bx$;$\btheta_k$) 
and parameters $\btheta_{k}$:
\begin{equation}
F(\bx;\bTheta)= \sum_{k=1}^{K}\pi_{k}f_{k}(\bx;\btheta_{k})%
\end{equation}
where $\pi_{k}$ are the component weights, constrained as $\sum_{k=1}^{K}\pi_{k}=1$. $\bTheta$ denotes the overall parameters of the mixture, 
estimated by maximizing the log-likelihood function:
\begin{equation}
\cL(\bTheta; \bx)= \log\prod_{i=1}^{N}F(\bx_i;\bTheta) = \sum_{i=1}^{N}\log \sum_{k=1}^{K}\pi_{k}f_{k}(\bx_i;\btheta_{k}) \, . \label{loglik}
\end{equation}
Usually the maximization has to be carried out numerically. However, for some particular choice of the components $f_k$, a closed form solution 
may be derived by using the \emph{Expectation-Maximization} (EM) algorithm. Notable examples are the multivariate normal and t-distribution and 
corresponding skew distributions, skew-normal and skew-t. In the EM framework the $N$ observed data are considered incomplete and a set of variable 
indicators $\bZ=(\bz_1,\ldots,\bz_N)$ is introduced, such that $\bz_i=(z_{1i},\ldots,z_{Ki})$ (i=1,\dots N) with $z_{ki}$=1 if $\bx_i$ 
comes from the $k$-th component and $z_{ki}$=0 otherwise.\\
Provided that the missing data are ignorable (MAR or MCAR), the missing features $\bx_{i,m}$ can be incorporated in the 
model as additional latent parameters together with the $z_{i}$.
In the following paragraphs we report the results obtained for normal and skew-normal mixtures.

\subsubsection{Normal Mixture Models}\label{NormalImputationSection}
Consider a multivariate normal (MN) distribution $f_{k}(\bx;\btheta_{k})$ as $k$-th component:
\begin{equation*}
f_{k}(\bx;\btheta_{k})= \frac{1}{2\pi |\bSigma_{k}|^{1/2}}\exp\left[-\frac{1}{2}(\bx-\bmu_{k})^{T}\bSigma_{k}^{-1}(\bx-\bmu_{k})\right]
\end{equation*}
with mean vector $\bmu_{k}$ and covariance matrix $\bSigma_{k}$ for the $k$-th component.\\ 
For the $i$-th observation the mixture parameters can be partitioned into an observed and missing part: ($\bmu_{ik,o}$, $\bmu_{ik,m}$),
($\bSigma_{ik,oo}$, $\bSigma_{ik,mm}$, $\bSigma_{ik,om}$).
Following \cite{Zoubin1994,Delalleau2012} the iterative solution provided by
the EM algorithm is given by:
\begin{align}
\tau_{ik}^{(t+1)}&= \frac{\pi_{k}^{(t)}f_{k}(\bx_{i,o},\btheta_{k,o}^{(t)})}{\sum_{k=1}^{K}\pi_{k}^{(t)}f_{k}(\bx_{i,o},\btheta_{k,o}^{(t)})}\\
\pi_{k}^{(t+1)}&= \frac{1}{N}\sum_{i=1}^{N}\tau_{ik}^{(t+1)}\\
\bmu_{k}^{(t+1)}&= \frac{\sum_{i=1}^{N}\tau_{ik}^{(t+1)}\hat{\bx}_{ik}^{(t+1)}}{\sum_{i=1}^{N}\tau_{ik}^{(t+1)}}\\%
\bSigma_{k}^{(t+1)}&= \frac{\sum_{i=1}^{N}\tau_{ik}^{(t+1)}(\hat{\bx}_{ik}^{(t+1)}-\bmu_{k}^{(t+1)})^{T}(\hat{\bx}_{ik}^{(t+1)}-
\bmu_{k}^{(t+1)})}{\sum_{i=1}^{N}\tau_{ik}^{(t+1)}}+\frac{\tau_{ik}^{(t+1)}\hat{\mathbf{C}}_{ik}^{(t+1)}}{\sum_{i=1}^{N}\tau_{ik}^{(t+1)}}
\end{align}
with $\hat{\bx}_{ik}$ and $\hat{\mathbf{C}}_{ik}$ respectively given by:
\begin{equation}
\hat{\bx}_{ik}=
\left\{
\begin{array}{l}
\hat{\bx}_{ik,o}= \bx_{i,o}\\%
\hat{\bx}_{ik,m}= \bmu_{k,m}+\bSigma_{k,mo}\bSigma_{k,oo}^{-1}(\bx_{i,o}-\bmu_{k,o}) 
\end{array}
\right.\\
\end{equation}
\begin{equation}
\hat{\mathbf{C}}_{ik}=
\left.
\begin{pmatrix}
\bZero & \bZero\\%
\bZero & \hat{\mathbf{C}}_{ik,mm}
\end{pmatrix}
\right.
,\;\hat{\mathbf{C}}_{ik,mm}= \bSigma_{k,mm}-\bSigma_{k,mo}\bSigma_{k,oo}^{-1}\bSigma_{k,mo}^{T}
\end{equation}
In the above expressions, $\hat{\bx}_{ik,m}=\mathbb{E}[z_{ik}\bx_{i,m}|\bx_{i,o},\btheta_{k}]$ is the expectation of the missing 
values $\bx_{i,m}$ given the observed data $\bx_{i,o}$, assuming that component $k$ generated $\bx_{i}$, while $\hat{\mathbf{C}}_{ik,mm}$ 
represents the covariance of the missing values $\bx_{i,m}$.\\
At convergence, the missing values $\bx_{i,m}$ are estimated by the weighted sum of $\hat{\bx}_{ik,m}$:
\begin{equation}
\bx_{i,m}= \sum_{k=1}^{K}\tau_{ik}\hat{\bx}_{ik,m}
\end{equation}
We note that if the $p$ variables $\bx$ are independent, the covariance matrix is diagonal and therefore $\bSigma_{k,mo}$=$\bZero$. In this
case the imputation rule yields $\bx_{i,m}$= $\sum_{k=1}^{K}\tau_{ik}\bmu_{k,m}$.

\subsubsection{Skew-Normal Mixture Models}\label{SkewNormalImputationSection}
Consider a multivariate skew-normal (MSN) distribution $f_{k}(\bx,\btheta_{k})$ as $k$-th component:
\begin{equation*}
f_{k}(\bx;\btheta_{k})= 2\phi_{p}(\bx;\bxi_{k},\bOmega_{k})\Phi(\bdelta_{k}^{T}\bOmega_{k}^{-1}(\bx-\bxi_{k});0,1-\bdelta_{k}^{T}\bOmega_{k}^{-1}\bdelta_{k})
\end{equation*}\label{MSNDefinition}
with location vector $\bxi_{k}$, scale matrix $\bSigma_{k}$, skewness vector $\bdelta_{k}$ and $\bOmega_{k}=\bSigma_{k}+\bdelta_{k}\bdelta_{k}^{T}$ for the $k$-th
component \footnote{Note that $\bxi$ and $\bSigma$ do not represent the mean and covariance of the distribution. These are given by:
\begin{eqnarray*}
&\mathbb{E}(\bX)=\bxi+\sqrt{\frac{2}{\pi}}\\
&\mbox{cov}(\bX)=\bSigma+\biggl(1-\frac{2}{\pi}\biggl)\bdelta\bdeltaT
\end{eqnarray*}\label{SkewNormalHierarchical}
}.
$\phi_{p}(\cdot)$ denotes the p-variate normal \emph{pdf} and $\Phi(\cdot)$ the \emph{cdf} of the univariate normal. 
It is worth noting that several forms of the skew-normal distribution exists in literature, both restricted and unrestricted. A nice 
comparison between different definitions is reported in \cite{LeeMcLachlan2012}. In this work we assumed the 
restricted form provided by Pyne et al \cite{Pyne2009}.\\
For the $i$-th observation the mixture parameters can be divided into an observed and missing part: ($\bxi_{ik,o}$, $\bxi_{ik,m}$),
($\bSigma_{ik,oo}$, $\bSigma_{ik,mm}$, $\bSigma_{ik,om}$), ($\bdelta_{ik,o}$,$\bdelta_{ik,m}$). 
A closed form solution to the problem has been provided by Lin et al \cite{Lin2009,Lin2010} 
for the unrestricted skew-normal model. In this work we explicitly derived the EM solution for the restricted form, given by: 
\begin{align*}
&\tau_{ik}^{(t+1)}= \frac{\pi_{k}^{(t)}f(\bx_{i,o},\btheta_{k,o}^{(t)})}{\sum_{k=1}^{K}\pi_{k}^{(t)}f_{k}(\bx_{i,o},\btheta_{k,o}^{(t)})}\\%
&\pi_{k}^{(t+1)}= \frac{1}{N}\sum_{i=1}^{N}\tau_{ik}^{(t+1)}\\%
&\bxi_{k}^{(t+1)}= \frac{\sum_{i=1}^{N}\tau_{ik}^{(t+1)}(\hat{\bx}_{ik}-\bdelta_{k}^{(t)}e_{1,ik}^{(t)})}{\sum_{i=1}^{N}\tau_{ik}^{(t+1)}}\\%
&\bdelta_{k}^{(t+1)}= \frac{\sum_{i=1}^{N}\tau_{ik}^{(t+1)}e_{1,ik}^{(t)}(\tilde{\bx}_{ik}-\bxi_{k}^{(t+1)})}{\sum_{i=1}^{N}\tau_{ik}^{(t+1)}e_{2,ik}^{(t)}}\\%
&\bSigma_{k}^{(t+1)}= \frac{\sum_{i=1}^{N}\tau_{ik}^{(t+1)}  (\hat{\bx}_{ik}-\bxi_{k}^{(t+1)}) (\hat{\bx}_{ik}-\bxi_{k}^{(t+1)})^{T} } 
{\sum_{i=1}^{N}\tau_{ik}^{(t+1)}} - 
\frac{\sum_{i=1}^{N}\tau_{ik}^{(t+1)} e_{1,ik}^{(t)}[\bdelta_{k}^{(t)}(\tilde{\bx}_{ik}-\bxi_{k}^{(t+1)})^{T}+(\tilde{\bx}_{ik}-\bxi_{k}^{(t+1)})
\bdelta_{k}^{T (t+1)}]}
{\sum_{i=1}^{N}\tau_{ik}^{(t+1)}}+\\%
+& \frac{\sum_{i=1}^{N}\tau_{ik}^{(t+1)} e_{2,ik}\bdelta_{k}^{(t+1)}\bdelta_{k}^{T (t+1)}]}
{\sum_{i=1}^{N}\tau_{ik}^{(t+1)}}+
\frac{\sum_{i=1}^{N}\tau_{ik}^{(t+1)}\hat{\mathbf{C}}_{ik}}{\sum_{i=1}^{N}\tau_{ik}^{(t+1)}}
\end{align*}
in which the needed expectation terms are given by:
\begin{align*}
&e_{1,ik}= \mu_{ik,o}+\sigma_{ik,o}\frac{\phi(\mu_{ik,o}/\sigma_{ik,o})}{\Phi(\mu_{ik,o}/\sigma_{ik,o})}\\%
&e_{2,ik}= \mu_{ik,o}^{2}+\sigma_{ik,o}^{2}+\mu_{ik,o}\sigma_{ik,o}
\frac{\phi(\mu_{ik,o}/\sigma_{ik,o})}{\Phi(\mu_{ik,o}/\sigma_{ik,o})}\\%
&\hat{\bx}_{ik}=
\left\{
\begin{array}{l}
\hat{\bx}_{ik,o}= \bx_{i,o}\\%
\hat{\bx}_{ik,m}= \bxi_{ik,m}+\bdelta_{ik,m}e_{1,ik}+\bSigma_{ik,mo}\bSigma_{ik,oo}^{-1}(\bx_{i,o}-\bxi_{ik,o}-
\bdelta_{ik,o}e_{1,ik})
\end{array}
\right.\\
&\tilde{\bx}_{ik}=
\left\{
\begin{array}{l}
\tilde{\bx}_{ik,o}= \bx_{i,o}\\%
\tilde{\bx}_{ik,m}= \bxi_{ik,m}+\bdelta_{ik,m}e_{2,ik}/e_{1,ik}+
\bSigma_{ik,mo}\bSigma_{ik,oo}^{-1}(\bx_{i,o}-\bxi_{ik,o}-\bdelta_{ik,o}e_{2,ik}/e_{1,ik})
\end{array}
\right.\\
&\hat{\mathbf{C}}_{ik}=
\left.
\begin{pmatrix}
\bZero & \bZero\\%
\bZero & \hat{\mathbf{C}}_{ik,mm}
\end{pmatrix}
\right.
,\;\hat{\mathbf{C}}_{ik,mm}=\bSigma_{ik,mm}-\bSigma_{ik,mo}\bSigma_{ik,oo}^{-1}\bSigma_{ik,om} + 
(e_{2,ik}-e_{1,ik}^2)(\bdelta_{ik,m}-\bSigma_{ik,mo}\bSigma_{ik,oo}^{-1}\bdelta_{ik,o})(\bdelta_{ik,m}-\bSigma_{ik,mo}\bSigma_{ik,oo}^{-1}\bdelta_{ik,o})^{T}
\end{align*}
with $\mu_{ik,o}=\bdelta_{ik,o}^T\bOmega_{ik,oo}^{-1}(\bx_{i,o}-\bxi_{ik,o})$ and $\sigma_{ik,o}^{2}=1-\bdelta_{ik,o}^{T}\bOmega_{ik,oo}^{-1}\bdelta_{ik,o}$.\\
In the above expressions, $\hat{\bx}_{ik,m}$ is the expectation of the missing 
values $\bx_{i,m}$ given the observed data $\bx_{i,o}$, assuming that component $k$ generated $\bx_{i}$, while $\hat{\mathbf{C}}_{ik,mm}$ 
represents the covariance of the missing values $\bx_{i,m}$.\\
At convergence, the missing values $\bx_{i,m}$ are estimated by the weighted sum of $\hat{\bx}_{ik,m}$:
\begin{equation}
\bx_{i,m}= \sum_{k=1}^{K}\tau_{ik}\hat{\bx}_{ik,m}
\end{equation}
More details on the derivation are reported in the \ref{SkewNormalAlgorithmDerivation} .

\section{Method Application}\label{MethodApplicationSection}
For our application we considered a test case, namely the interaction of charged particles (pions, kaons and protons) on a multilayer Silicon detector, considering
the energy losses in each layer as relevant observables. This is a well known technique in particle and nuclear physics 
to separate and identify different particles according to their 
specific energy loss as a function of the particle momentum, usually provided by their curvature in a magnetic field or by the time-of-flight
measurement.
Missing data may arise in this scenario either from detector inefficiencies, i.e. inactive or damaged detector areas, or
from faulty electronic channels.
Even though the number of assumed variables in this example is not large (six layers were considered, as is the case of a typical inner
tracking detector, such as the present configuration of the ALICE detector at LHC\cite{ALICE}), the case 
may be usefully employed to demonstrate the 
capabilities of the various approaches and compare their performance. 
For the classification of charged particles we adopted a neural network approach, comparing its performance according to the way 
in which missing data were previously imputed by one of the methods 
described in Section \ref{MissingDataSection}.

\begin{figure}[!t]
\centering
\includegraphics[scale=0.4]{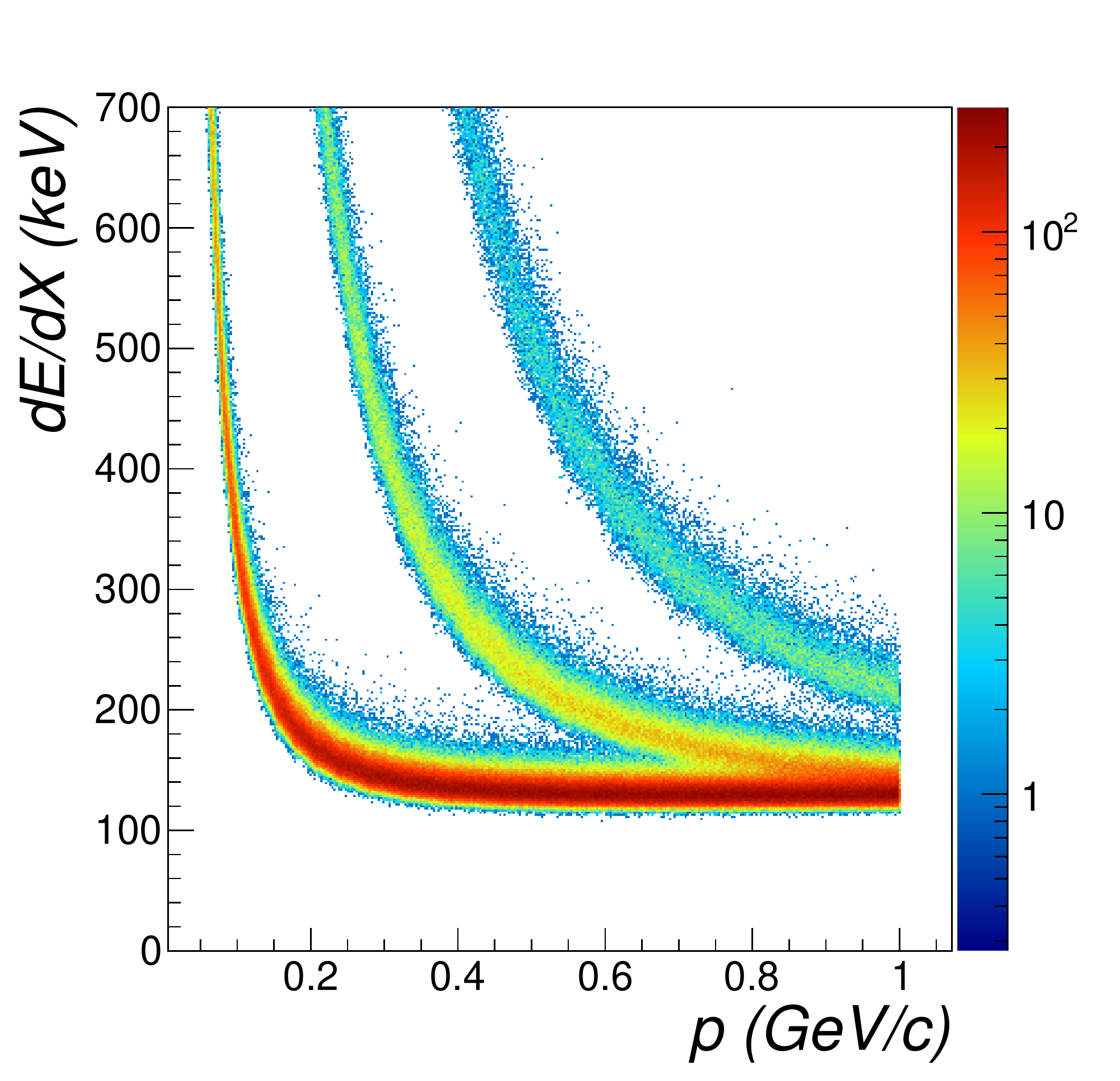}%
\vspace{0.1cm}
\caption{Specific energy loss $dE/dX$ for pions, kaons and protons as a function of their particle momentum, as obtained from a GEANT 
simulation. The average of the two smallest energy deposits in the six layers is considered in the plot.}
\label{EnergyLossVSMomentum}
\end{figure}

\begin{figure}[!h]
\centering
\includegraphics[scale=0.13]{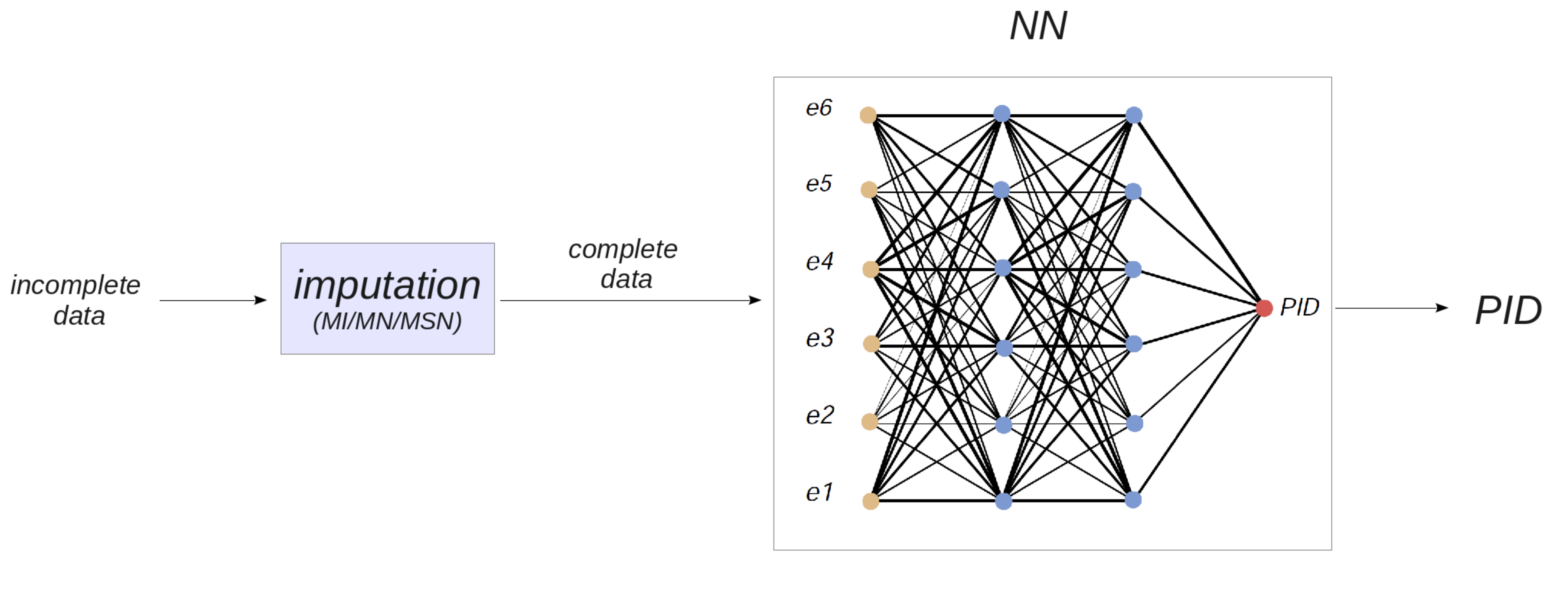}%
\vspace{0.1cm}
\caption{Schema of the analysis method.}
\label{AnalysisMethodSketchFig}
\end{figure}

\subsection{Simulated data set}\label{SimulationSection}
To generate the simulated data set, the interaction of $\pi^+$, $K^+$ and protons was implemented in a GEANT software replica of a six-layers 
Silicon plane detector. Each layer had 500 $\mu$m thickness.
Particles were generated with a relative abundance of 0.80/0.15/0.05 for pions/kaons/protons, with a uniform momentum distribution in the 
interval 0-1 GeV/c. 
The energy loss deposited in each layer was stored event-by-event for subsequent analysis. About 10$^{6}$ events were produced. 
Since the identification performances are expected to strongly depends on the particle momentum, 
the data set was divided into individual momentum bins of 
width 50 MeV/c. For each bin, half of the data sample was used as training and cross validation for the 
neural network learning stage. 

\begin{figure}[!t]
\centering
\subtable[MN imputation]{\includegraphics[scale=0.4]{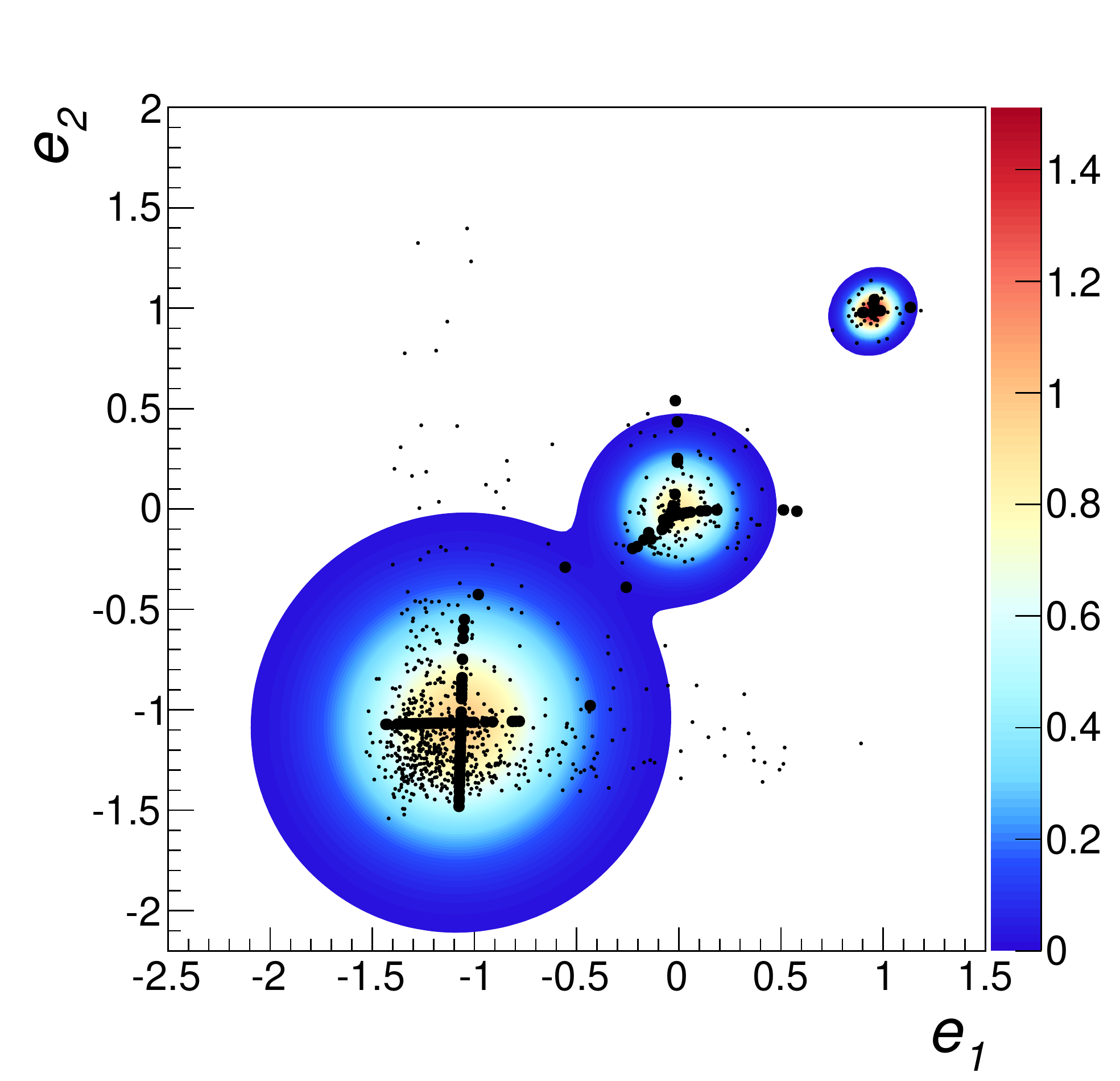}\label{SampleImputationPlot1}}%
\subtable[MSN imputation]{\includegraphics[scale=0.4]{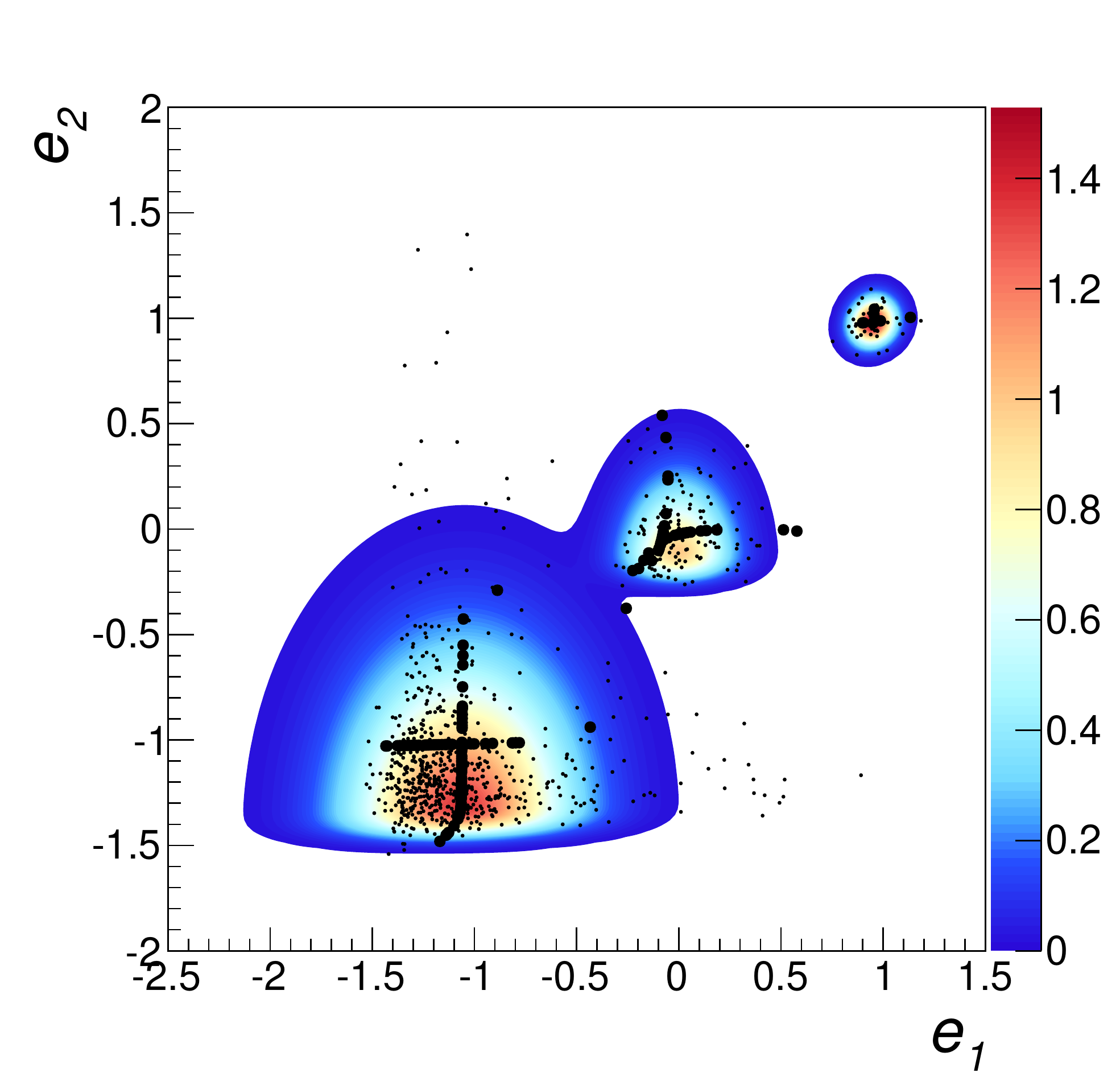}\label{SampleImputationPlot2}}%
\vspace{0.1cm}
\caption{Normal (left panel) and skew-normal (right panel) imputation results for a sample data set with 10\% missing data. 
True (simulated) data are shown in black filled dots, while imputed data with open black dots. The colored area represents the 
contour levels of the fitted normal/skew-normal mixture model.}
\label{SampleImputationPlot}
\end{figure}

The remaining part of the data set available in the momentum bin was used to draw $N_{\textsf{s}}$=100 random test 
samples with $N$=1000 events. We generated randomly missing data in each test sample according to a prespecified missing probability, 
assumed equal for 
each detector layer. Such samples were finally used to test the considered imputation methods.\\
Fig. \ref{EnergyLossVSMomentum} shows a typical scatter plot of the specific energy loss 
$dE/dX$ as a function of the particle momentum $p$. In particular we reported in the plot the truncated energy loss, computed by taking the average of the 
two smallest values in the six detector layers. 

\begin{figure}[!t]
\centering
\subtable[$p$=(0.25-0.30) GeV/c]{\includegraphics[scale=0.3]{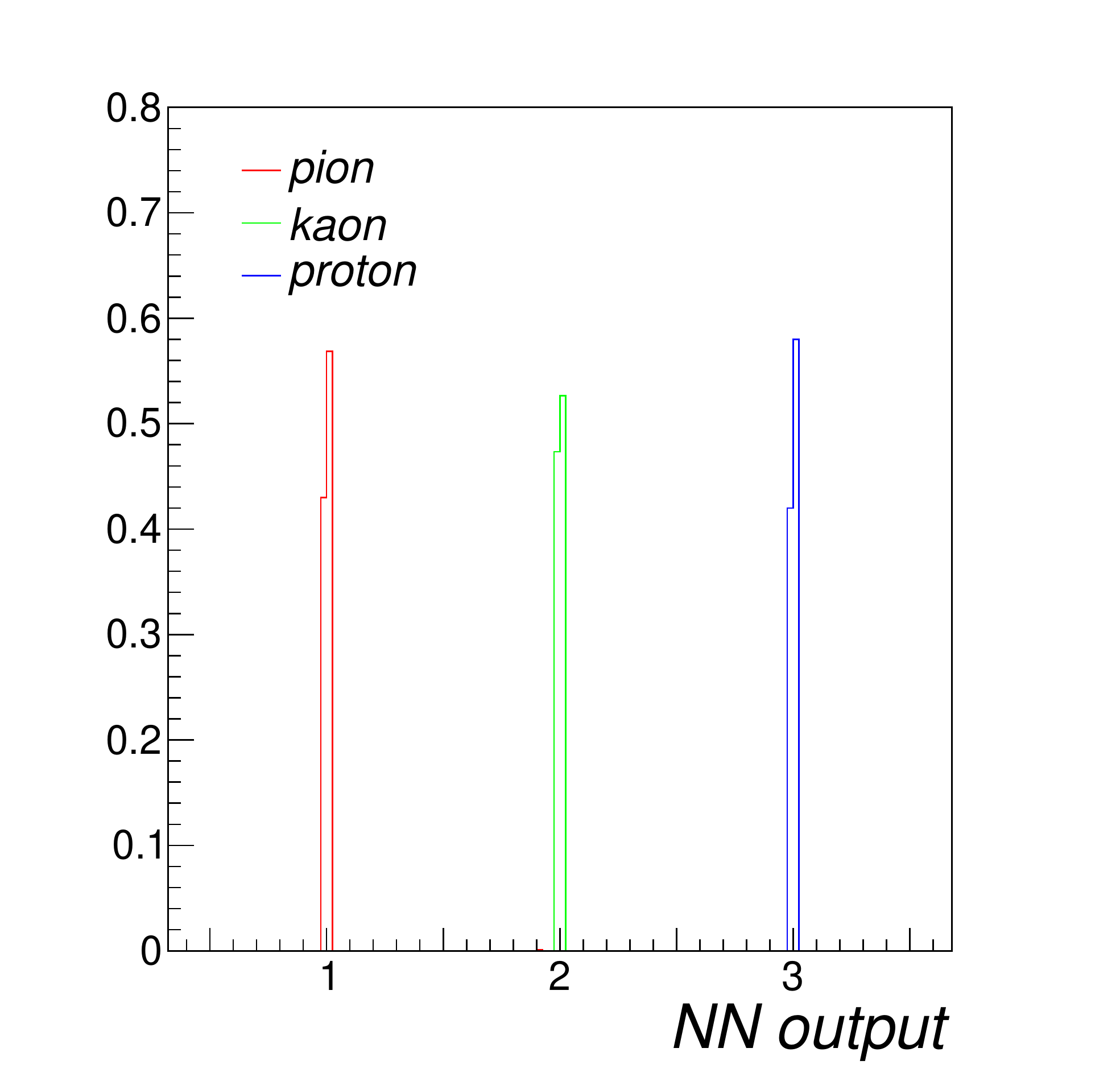}\label{NNOutputPlot1}}%
\hspace{-0.9cm}
\subtable[$p$=(0.55-0.60) GeV/c]{\includegraphics[scale=0.3]{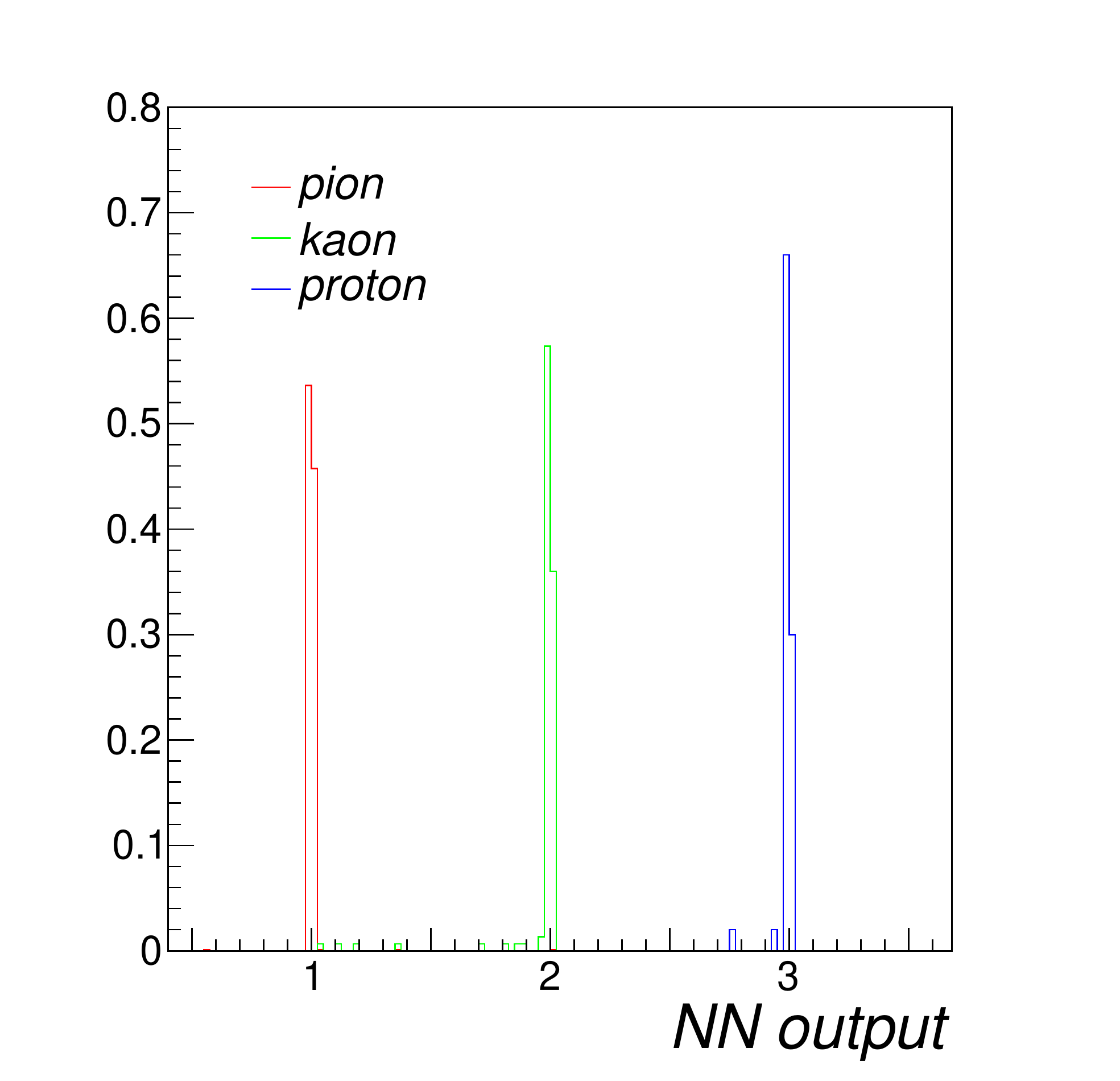}\label{NNOutputPlot2}}%
\hspace{-0.9cm}
\subtable[$p$=(0.85-0.90) GeV/c]{\includegraphics[scale=0.3]{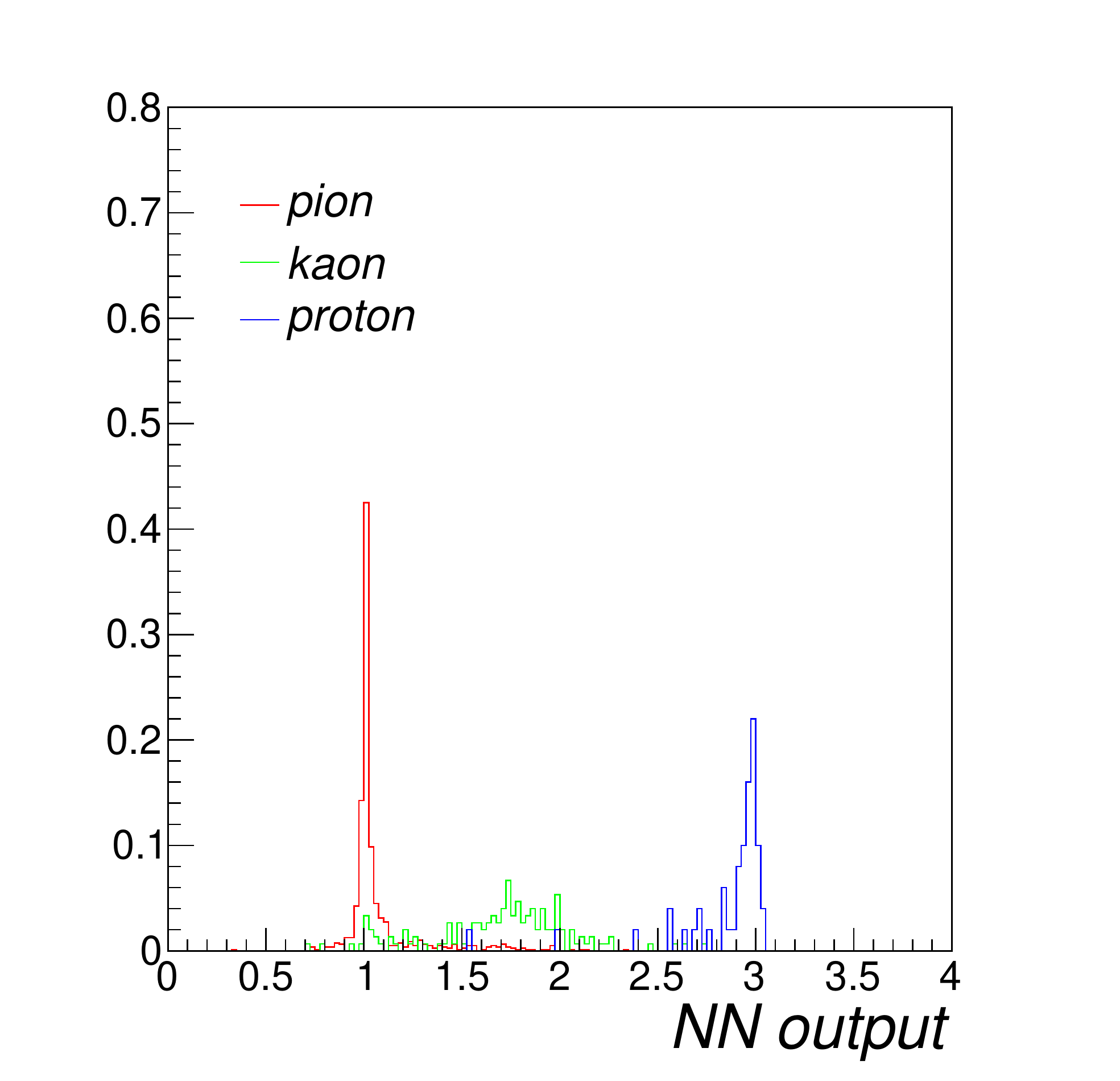}\label{NNOutputPlot3}}%
\vspace{0.1cm}
\caption{Distribution of the neural network outputs of pions(red histogram), kaons (green histogram) and protons (blue histogram) events 
obtained when presenting a sample data set to the designed network for three different momentum bins: 
(0.25-0.30) GeV/c (left panel), (0.55-0.60) GeV/c (central panel) and (0.85-0.90) GeV/c (right panel).}
\label{NNOutputPlot}
\end{figure}

The individual  energy loss distribution for a given momentum bin has the typical Landau shape, with an asymmetric tail at high energy loss. 
This aspect is important, since most of the available algorithms to impute missing data are most suited for Gaussian distributions.\\
To reduce the impact of Landau tails and get rid of the momentum dependence we normalized the simulated energy losses obtained in the 
six layers to the logarithm of the average energy loss for kaons, e.g. we considered $e\equiv\ln(dE/dX)-\ln(dE/dX_{kaons}(p))$ as variable 
for the classification analysis.

\subsection{Neural network approach to particle identification}\label{NeuralNetworkSection}
Neural network approaches to particle identification in nuclear and particle physics have been extensively used in the past. In this work we considered
a feed-forward neural network. The strategy is sketched in Fig. \ref{AnalysisMethodSketchFig}. 
We train a neural network for each momentum bin 
with a set of complete simulated data. After a few trials, a 6:6:6:1 network topology was chosen for the entire momentum range,
where the input neurons were the 
six energy losses in the different layers and the output layer is the particle identification (PID) code (1=pion, 2=kaon, 3=proton). 
Two hidden layers with six neurons each were introduced to improve the particle separation capabilities at high momenta. Hyperbolic tangent
activation functions
were used for the neurons in the hidden layer. The Broyden-Fletcher-Goldfarb-Shanno (BFGS) algorithm was employed as learning method and a number of 
iterations in the order of 1000 provides both optimal classification results and network generalization capabilities.\\The trained network is then used to 
identify the test data samples in which missing data were previously imputed separately with each of the methods presented in the 
previous Section: Mean, 
Maximum Likelihood for MN and MSN mixtures and Multiple Imputation. The maximum-likelihood algorithms were completely developed in C++ employing 
the implementation of multivariate distributions of the R statistical tool \cite{RProject} via the RInside/Rcpp interface \cite{RInside}.
Several implementation to perform multiple imputation are available in current statistical tools. 
To compare with other algorithms we made use for this work of the \textsc{Amelia II} package \cite{AmeliaPackage} present in the R tool.

\begin{figure}[!t]
\centering
\subtable[$p$=(0.25-0.30) GeV/c]{\includegraphics[scale=0.27]{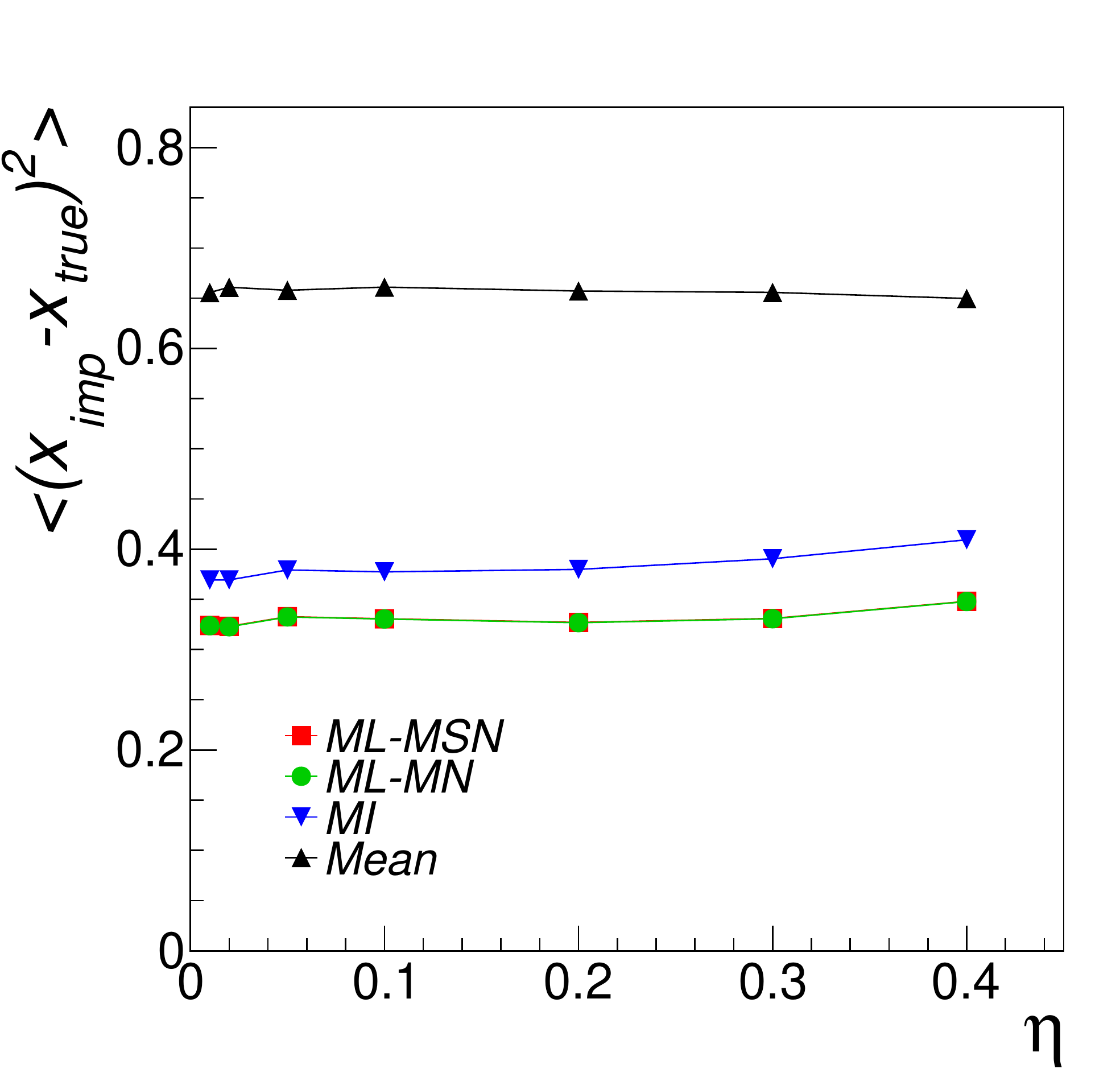}\label{DiffPlot1}}%
\hspace{-0.25cm}
\subtable[$p$=(0.55-0.60) GeV/c]{\includegraphics[scale=0.27]{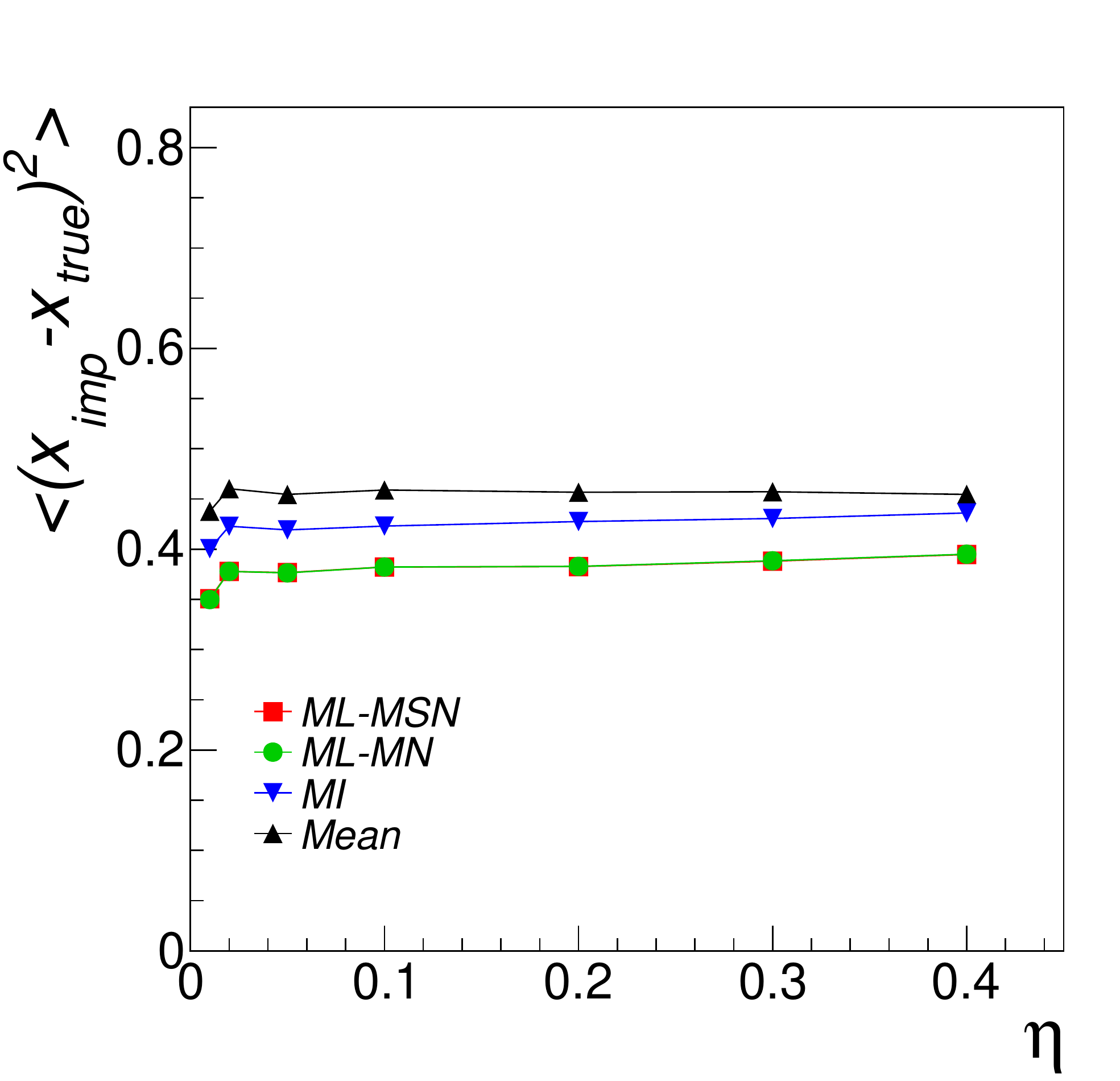}\label{DiffPlot2}}%
\hspace{-0.25cm}
\subtable[$p$=(0.85-0.90) GeV/c]{\includegraphics[scale=0.27]{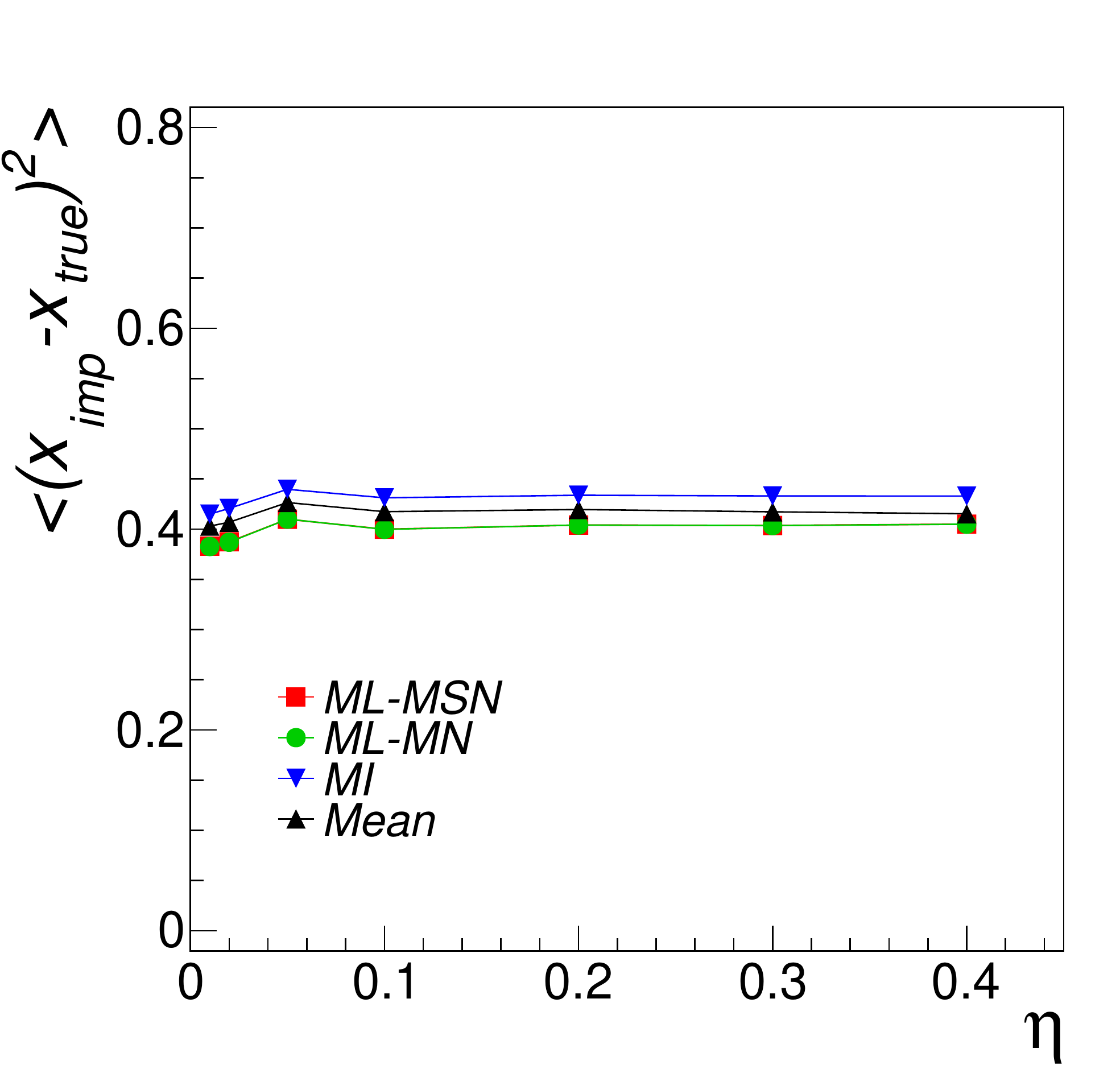}\label{DiffPlot3}}%
\vspace{0.1cm}
\caption{Average differences between the true data and the imputed data, 
for three different momentum bins: 
(0.25-0.30) GeV/c (left panel), (0.55-0.60) GeV/c (central panel) and (0.85-0.90) GeV/c (right panel).}
\label{Differences}
\end{figure}

\begin{figure}[!th]
\centering
\subtable[$p$=(0.25-0.30) GeV/c - pion]{\includegraphics[scale=0.28]{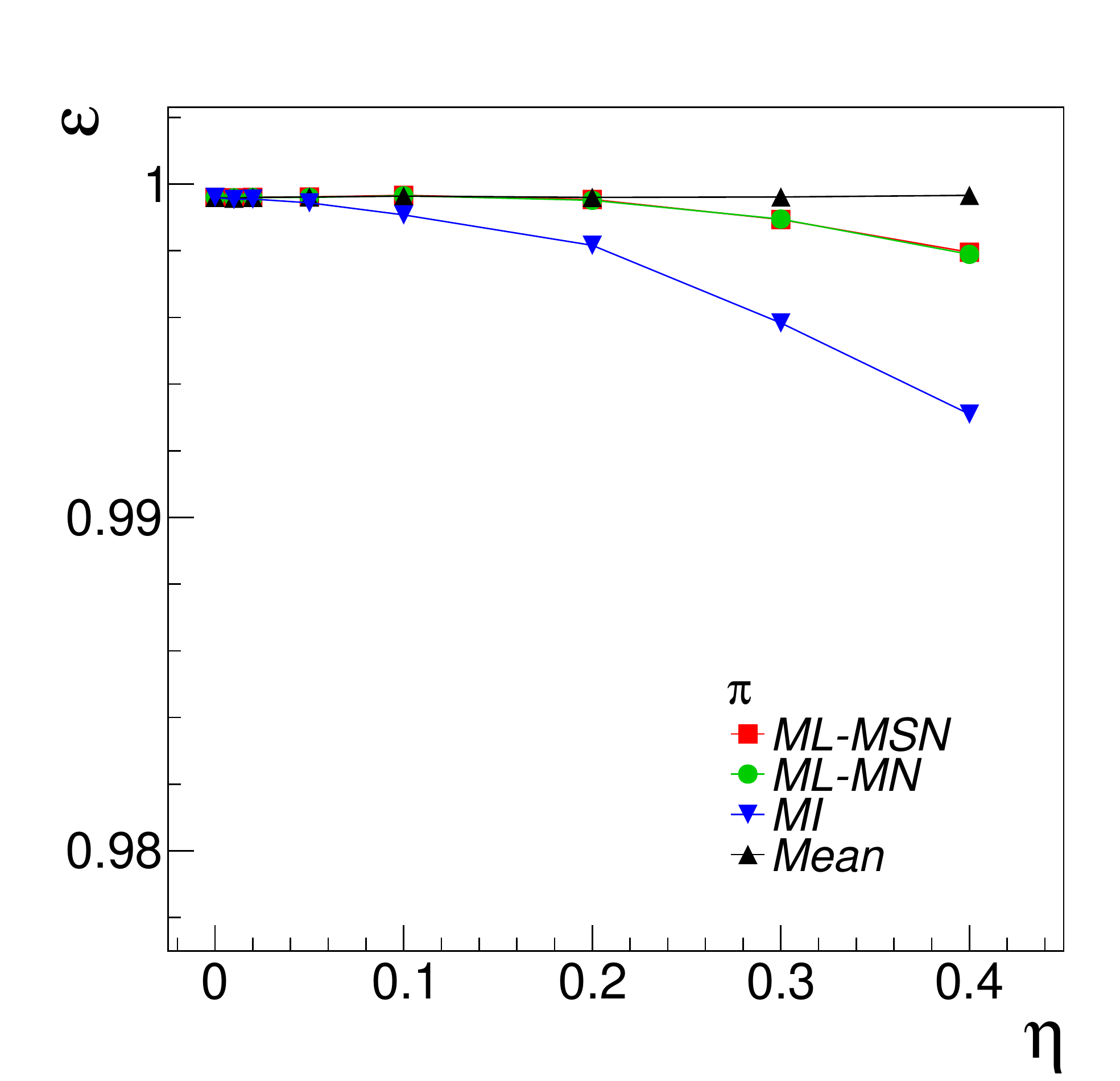}\label{EfficiencyLowMomentumPlot1}}%
\hspace{-0.2cm}%
\subtable[$p$=(0.25-0.30) GeV/c - kaon]{\includegraphics[scale=0.28]{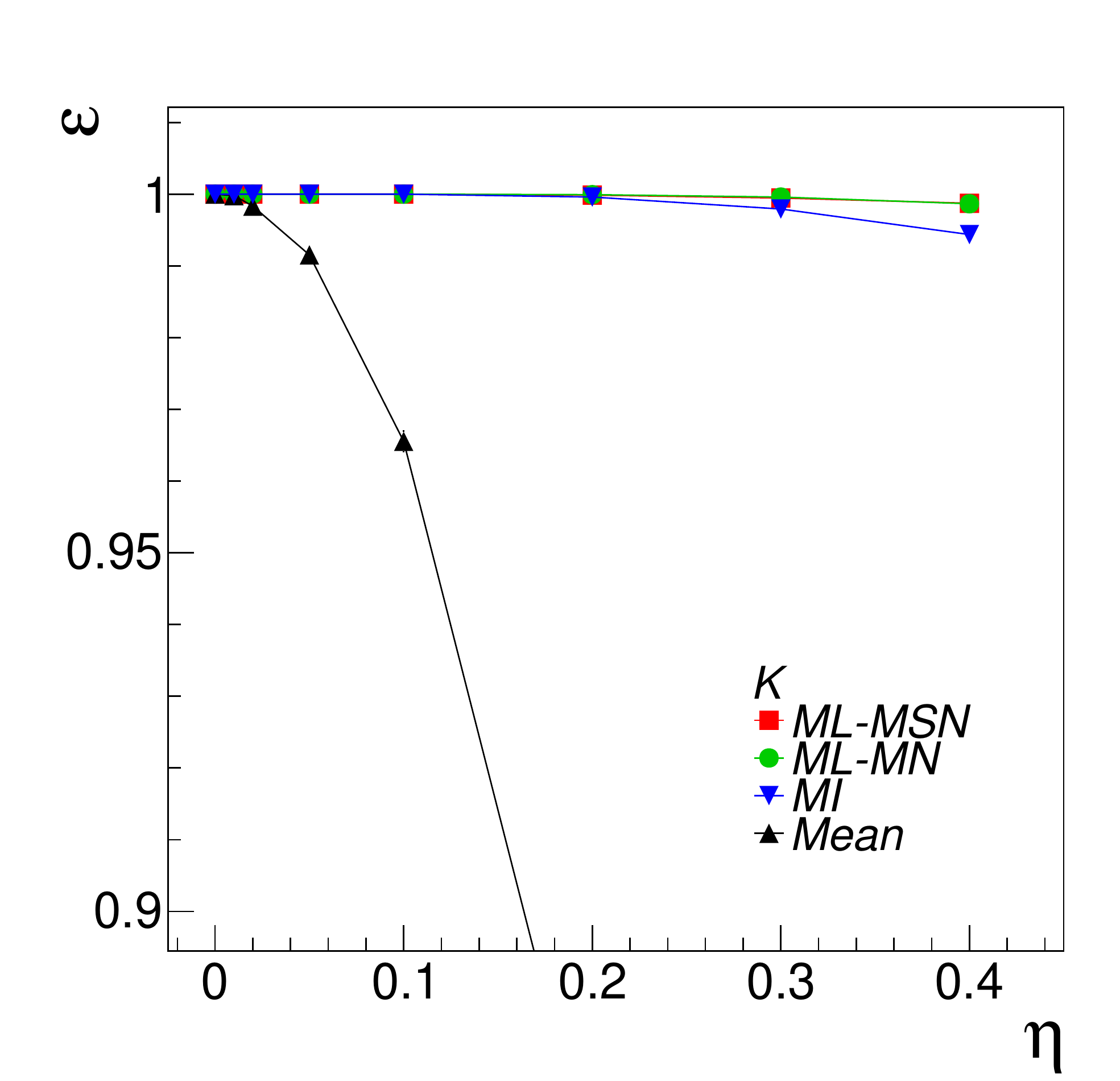}\label{EfficiencyLowMomentumPlot2}}%
\hspace{-0.2cm}%
\subtable[$p$=(0.25-0.30) GeV/c - proton]{\includegraphics[scale=0.28]{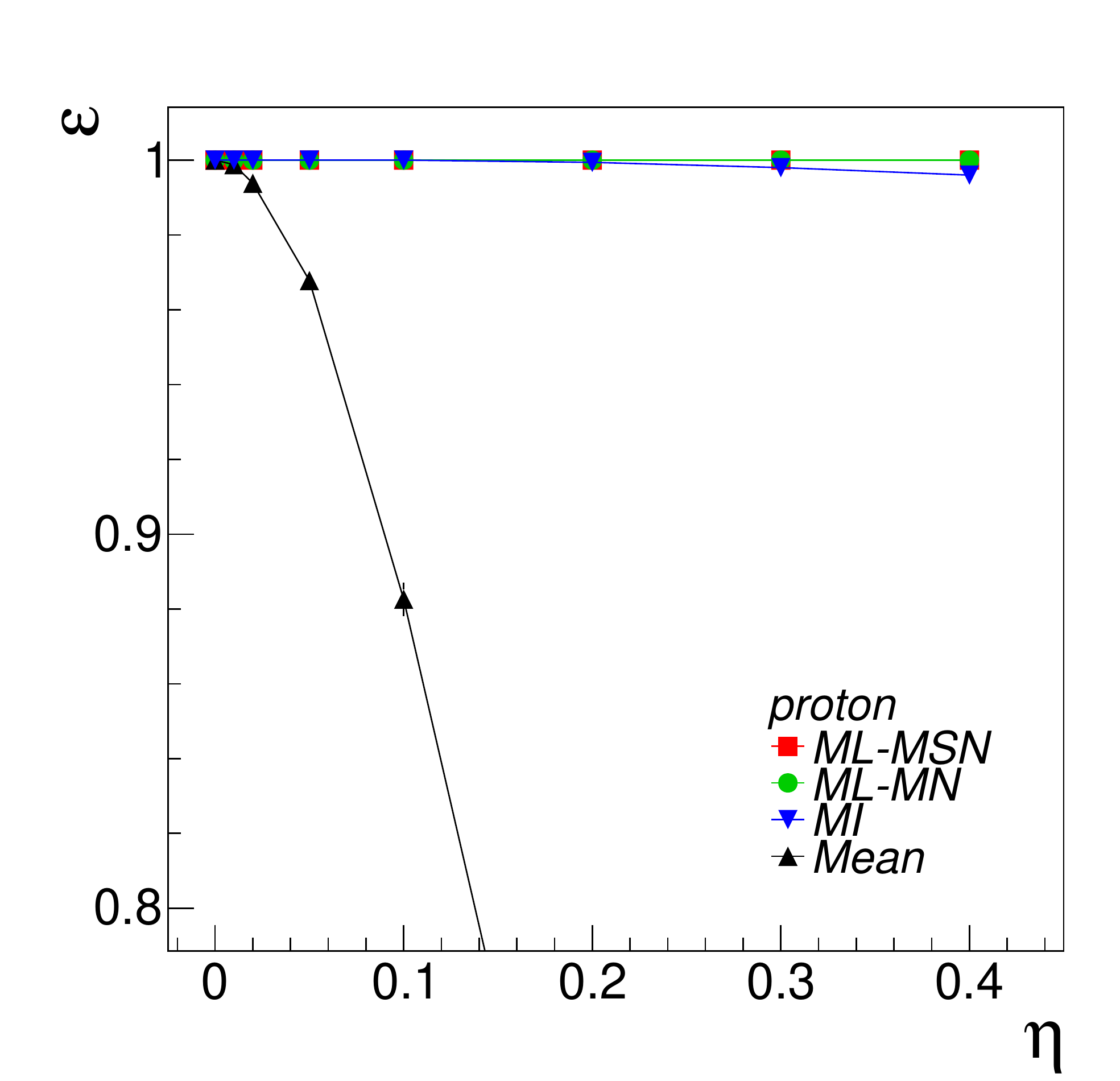}\label{EfficiencyLowMomentumPlot3}}%
\\%
\vspace{-0.3cm}
\subtable[$p$=(0.55-0.60) GeV/c - pion]{\includegraphics[scale=0.28]{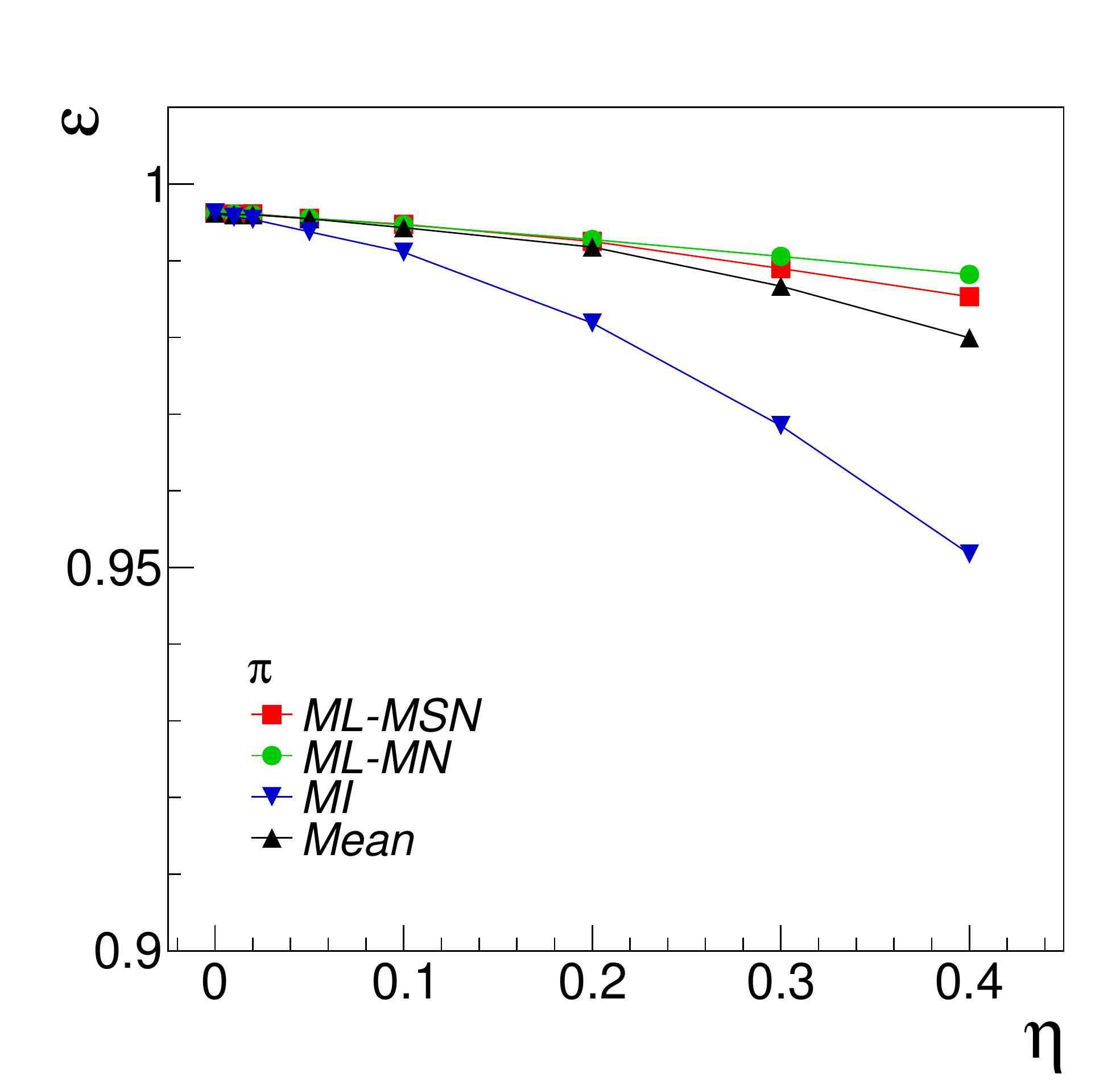}\label{EfficiencyMediumMomentumPlot1}}%
\hspace{-0.2cm}%
\subtable[$p$=(0.55-0.60) GeV/c - kaon]{\includegraphics[scale=0.28]{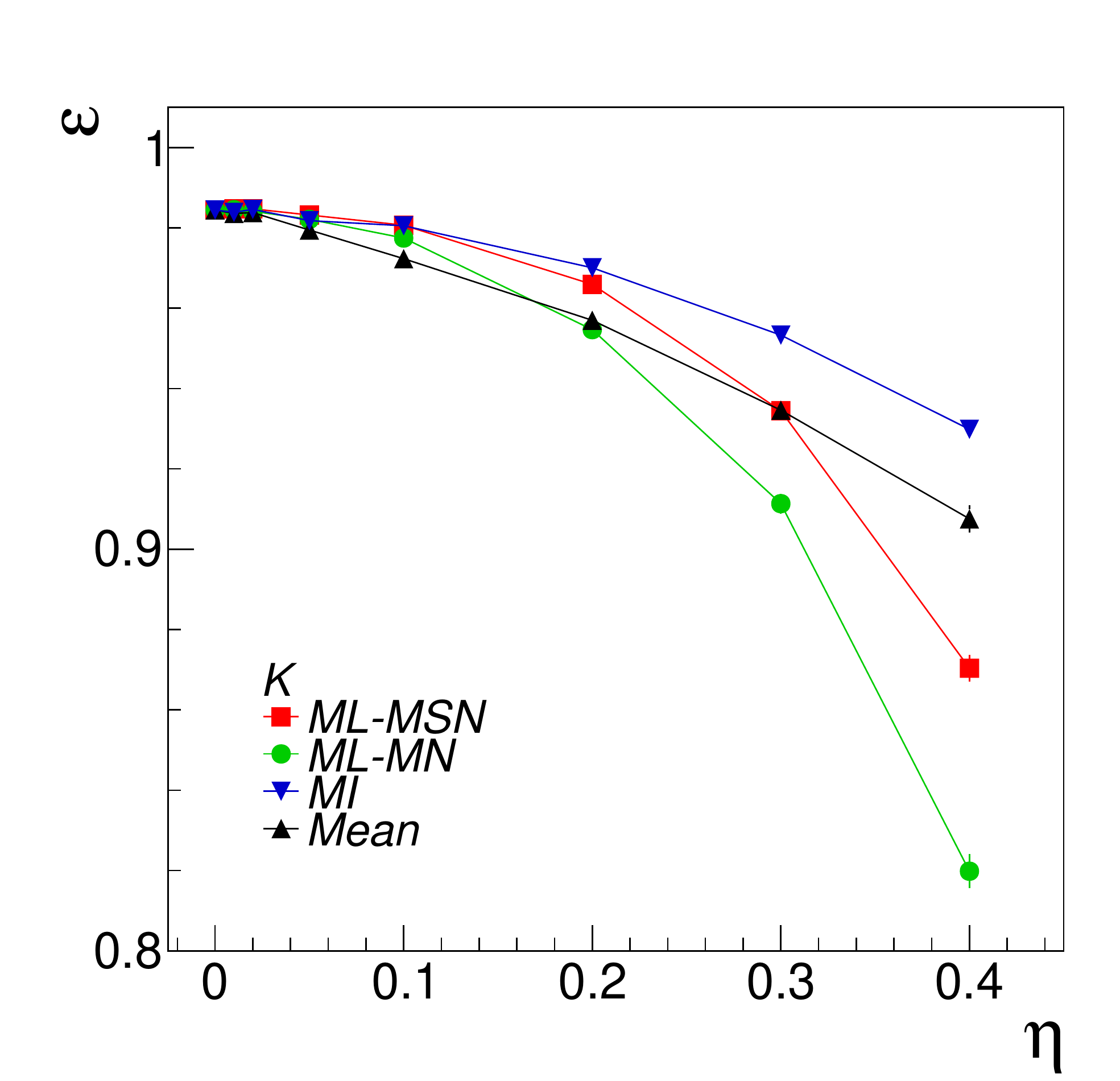}\label{EfficiencyMediumMomentumPlot2}}%
\hspace{-0.2cm}%
\subtable[$p$=(0.55-0.60) GeV/c - proton]{\includegraphics[scale=0.28]{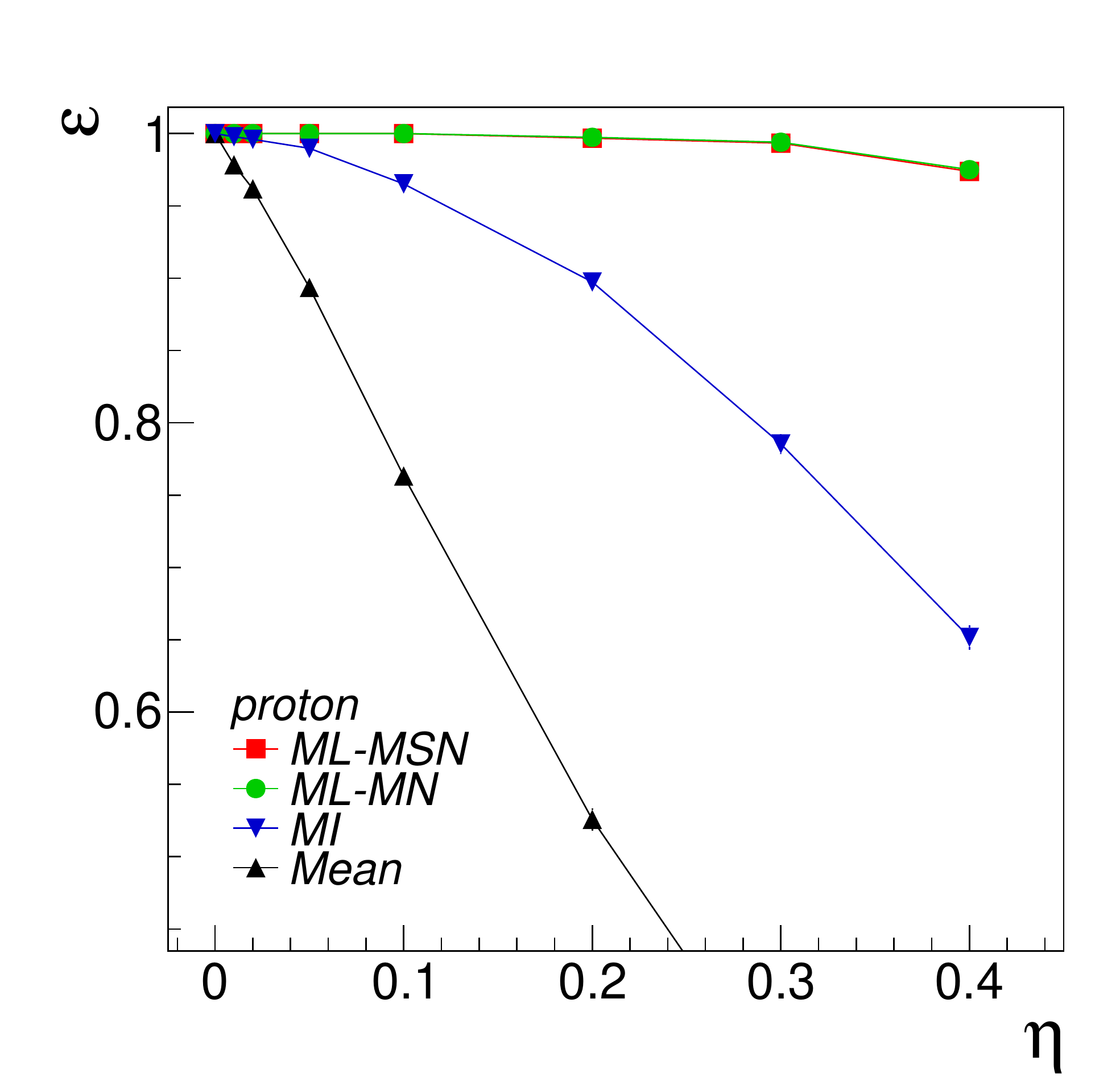}\label{EfficiencyMediumMomentumPlot3}}%
\\%
\vspace{-0.3cm}
\subtable[$p$=(0.85-0.90) GeV/c - pion]{\includegraphics[scale=0.28]{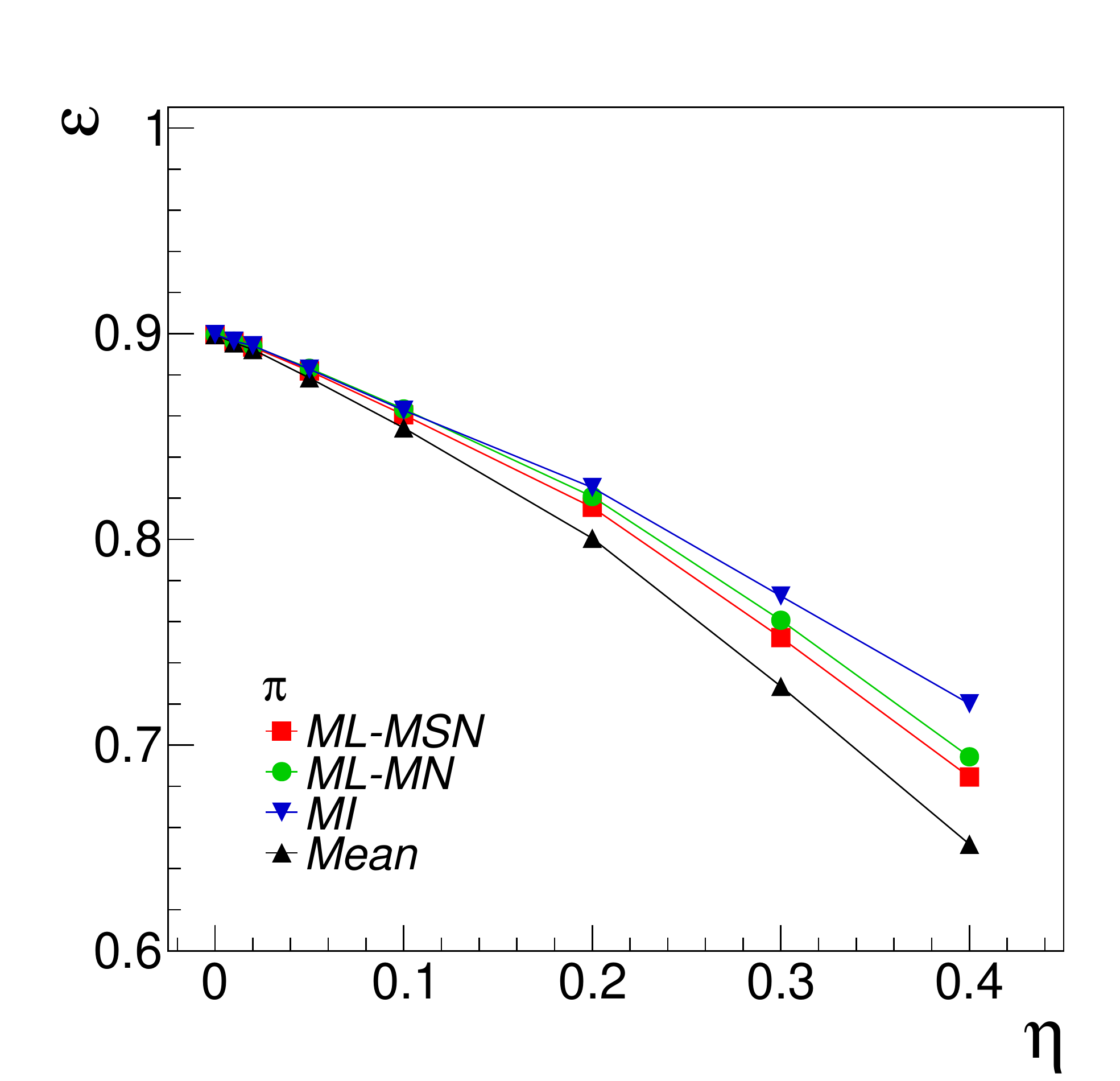}\label{EfficiencyHighMomentumPlot1}}%
\hspace{-0.2cm}%
\subtable[$p$=(0.85-0.90) GeV/c - kaon]{\includegraphics[scale=0.28]{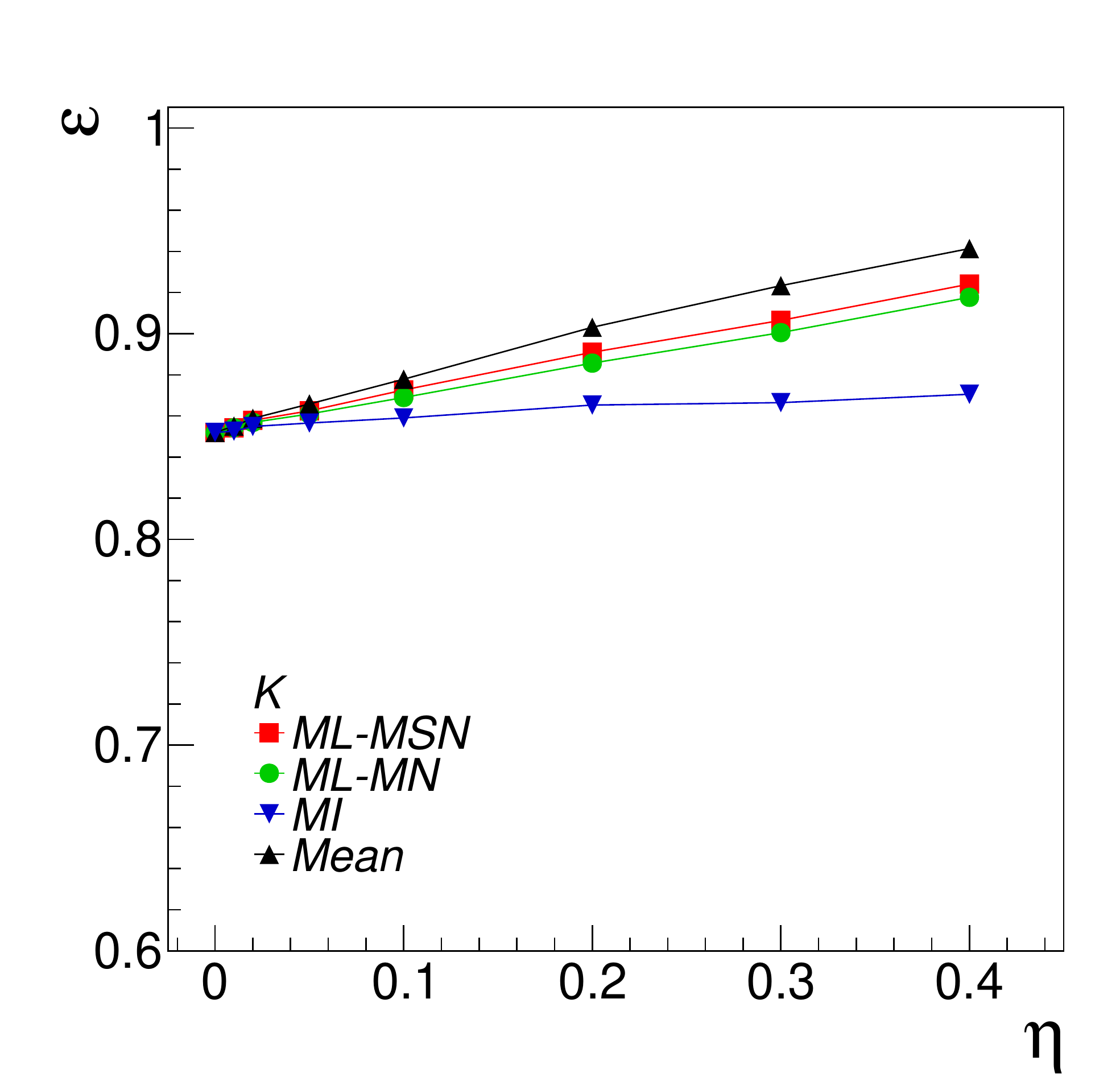}\label{EfficiencyHighMomentumPlot2}}%
\hspace{-0.2cm}%
\subtable[$p$=(0.85-0.90) GeV/c - proton]{\includegraphics[scale=0.28]{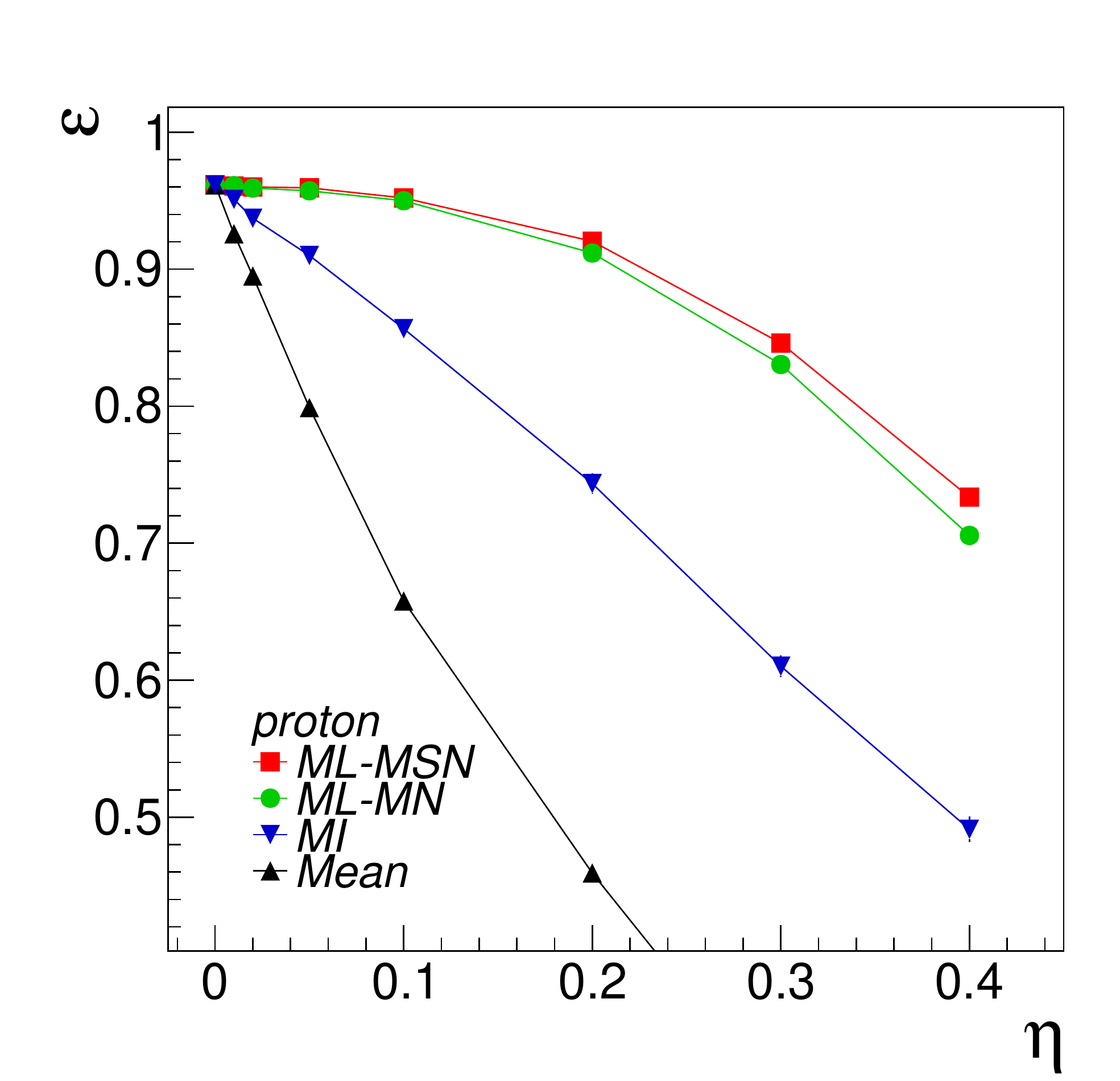}\label{EfficiencyHighMomentumPlot3}}
\vspace{-0.2cm}
\caption{Classification efficiency for pions, kaons and protons, respectively shown from left to right obtained with different imputation methods
for three  momentum bins: (0.25-0.3) GeV/c (top panels), (0.55-0.60) GeV/c (middle panels) and  (0.85-0.90) GeV/c (bottom panels). Multiple imputation (MI) results
are shown with blue triangle markers, mean imputation with black markers, maximum likelihood with multivariate normal model (ML-MN) with green dots 
and maximum likelihood with multivariate skew-normal model (ML-MSN) with red squares.}
\label{EfficiencyPlot}
\end{figure}

\begin{figure}[!th]
\centering
\subtable[$p$=(0.25-0.30) GeV/c - pion]{\includegraphics[scale=0.28]{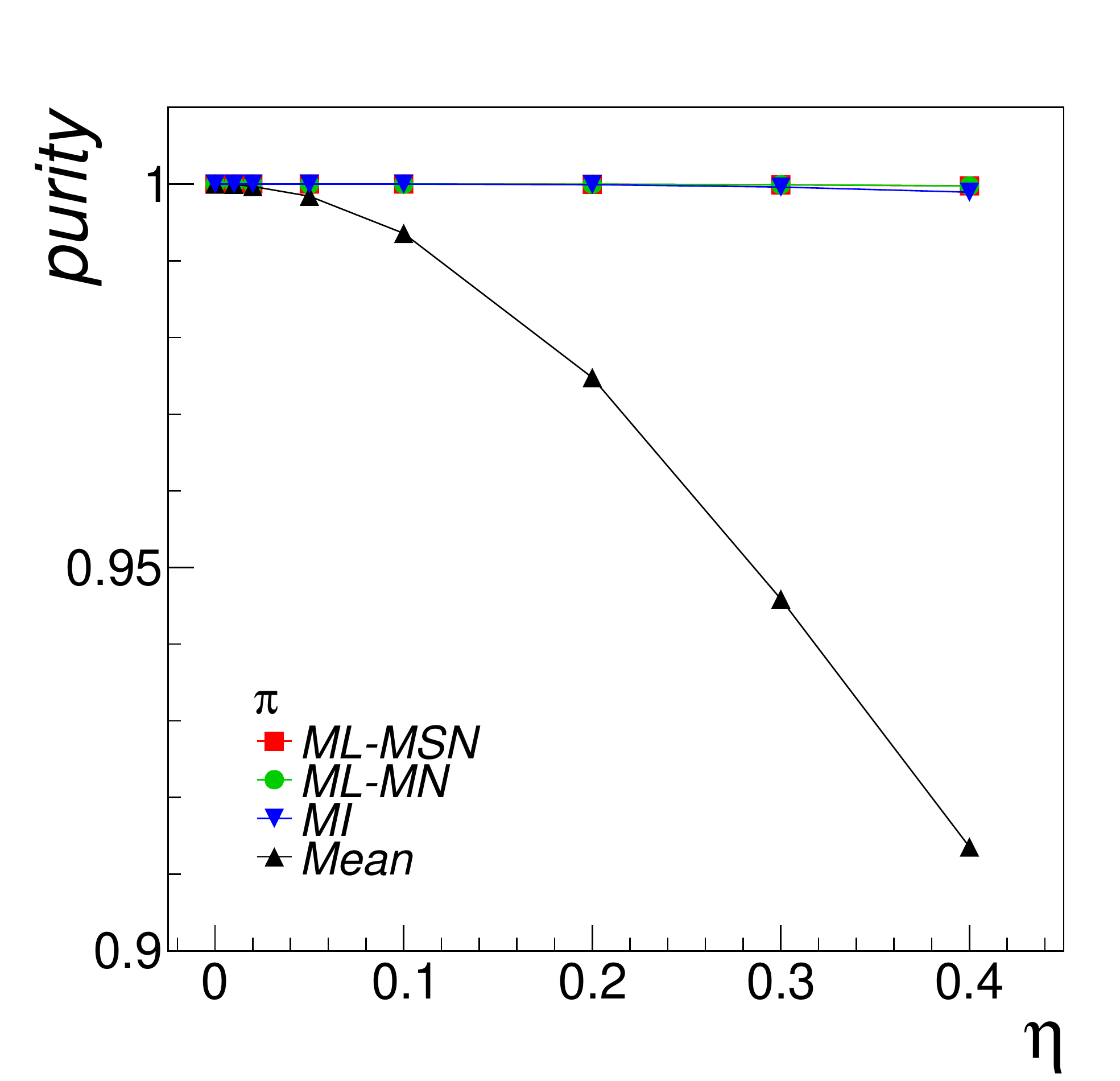}\label{PurityLowMomentumPlot1}}%
\hspace{-0.2cm}%
\subtable[$p$=(0.25-0.30) GeV/c - kaon]{\includegraphics[scale=0.28]{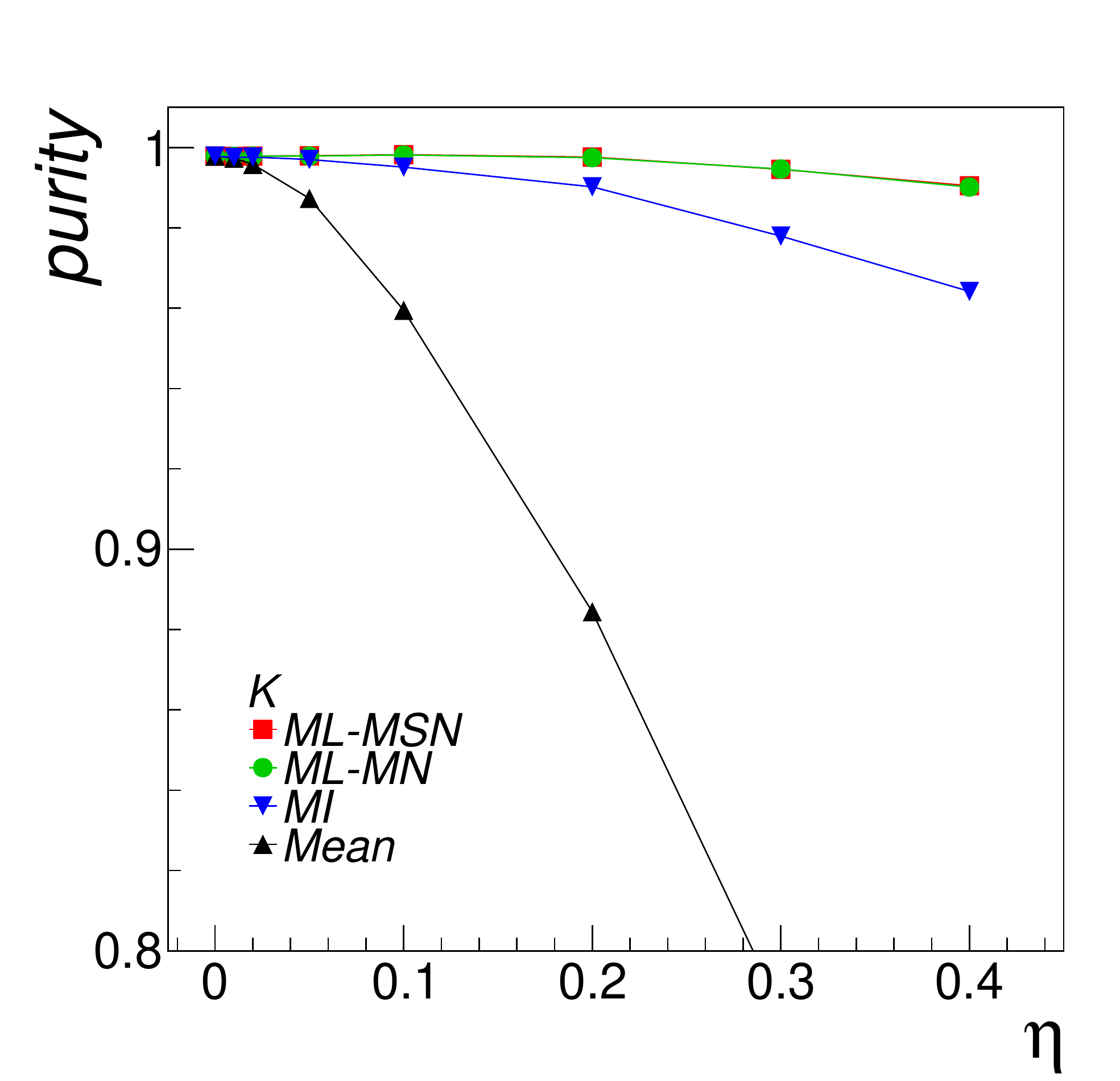}\label{PurityLowMomentumPlot2}}%
\hspace{-0.2cm}%
\subtable[$p$=(0.25-0.30) GeV/c - proton]{\includegraphics[scale=0.28]{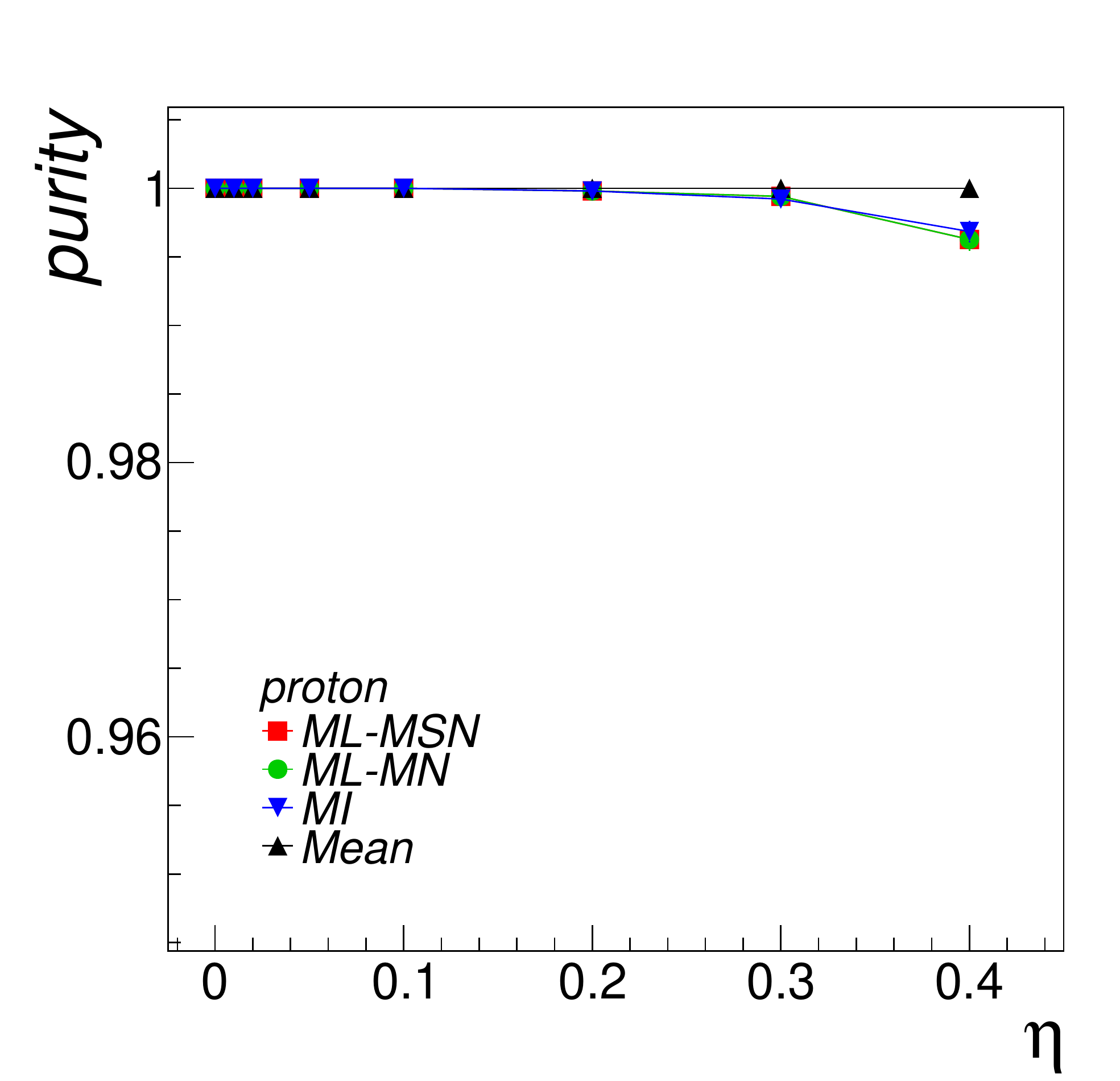}\label{PurityLowMomentumPlot3}}%
\\%
\vspace{-0.3cm}
\subtable[$p$=(0.55-0.60) GeV/c - pion]{\includegraphics[scale=0.28]{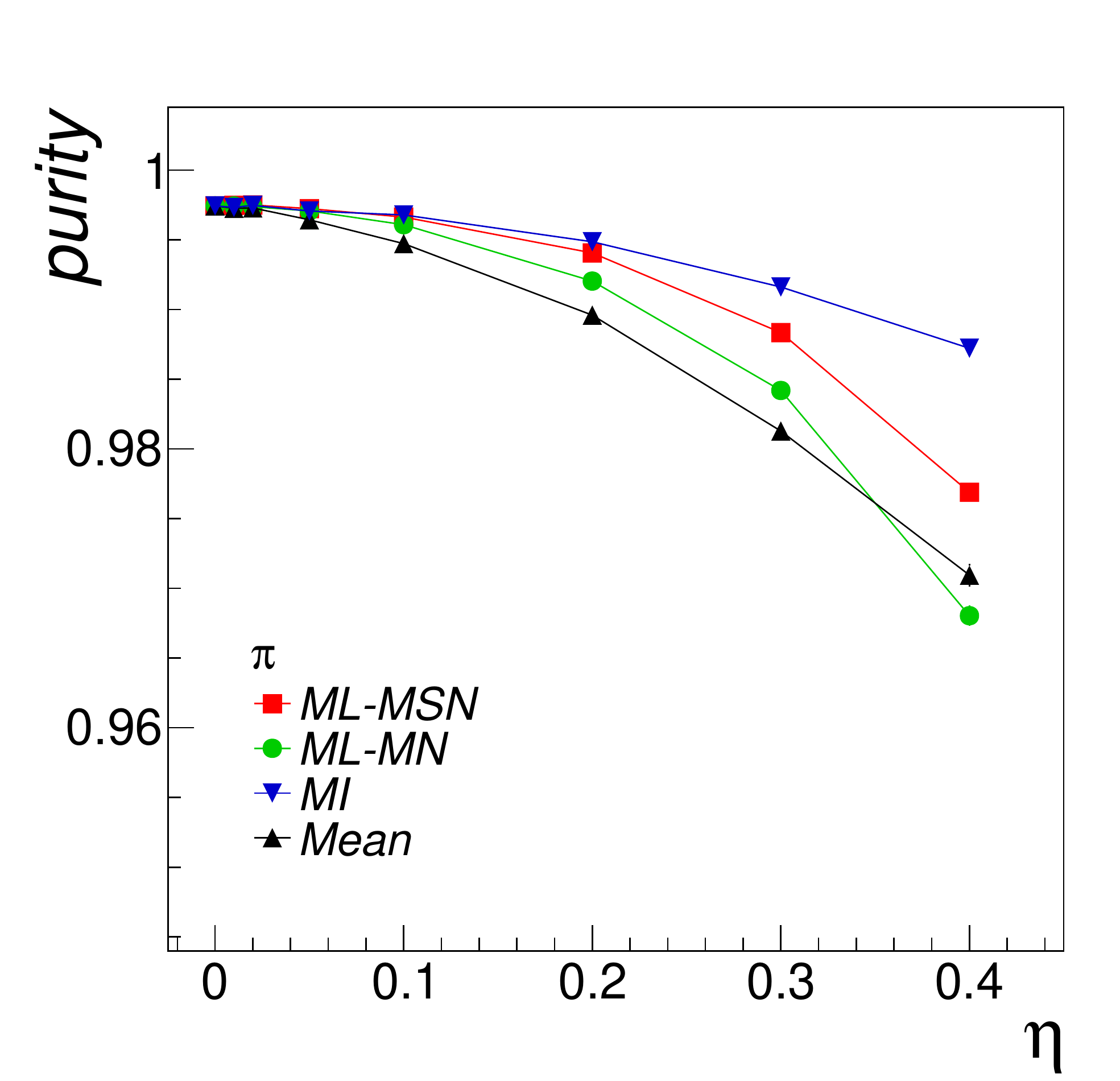}\label{PurityMediumMomentumPlot1}}%
\hspace{-0.2cm}%
\subtable[$p$=(0.55-0.60) GeV/c - kaon]{\includegraphics[scale=0.28]{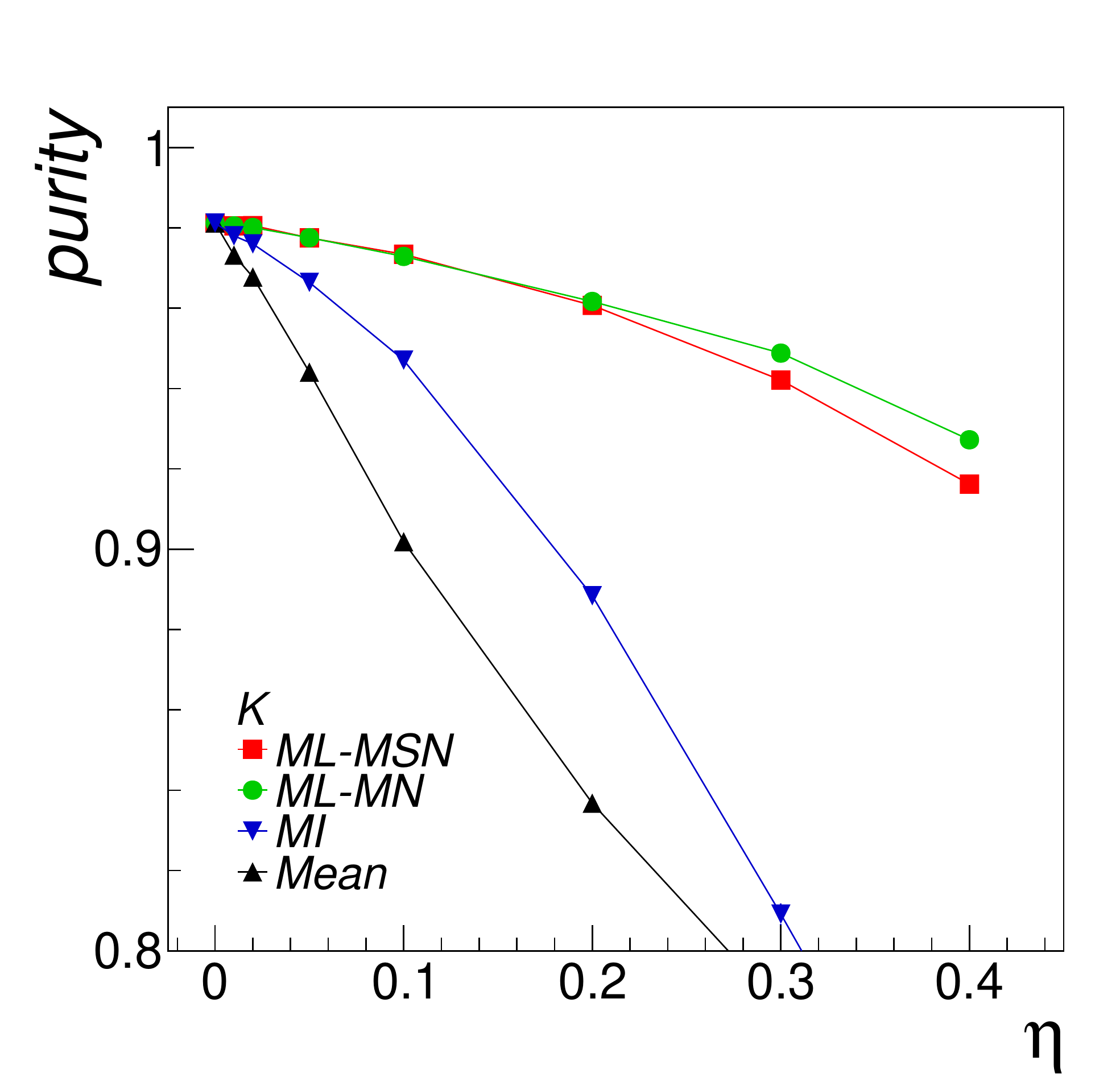}\label{PurityMediumMomentumPlot2}}%
\hspace{-0.2cm}%
\subtable[$p$=(0.55-0.60) GeV/c - proton]{\includegraphics[scale=0.28]{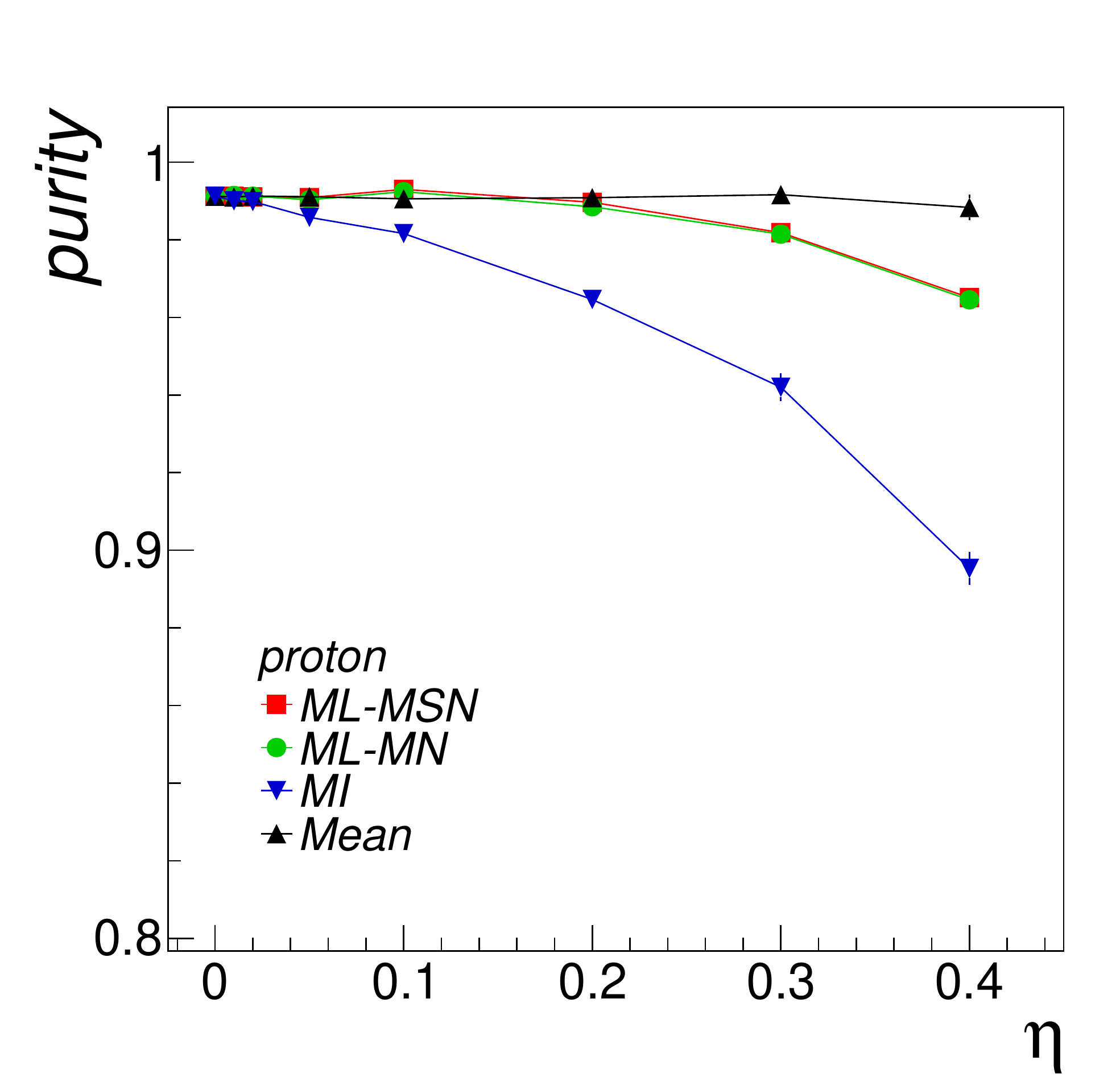}\label{PurityMediumMomentumPlot3}}%
\\%
\vspace{-0.3cm}
\subtable[$p$=(0.85-0.90) GeV/c - pion]{\includegraphics[scale=0.28]{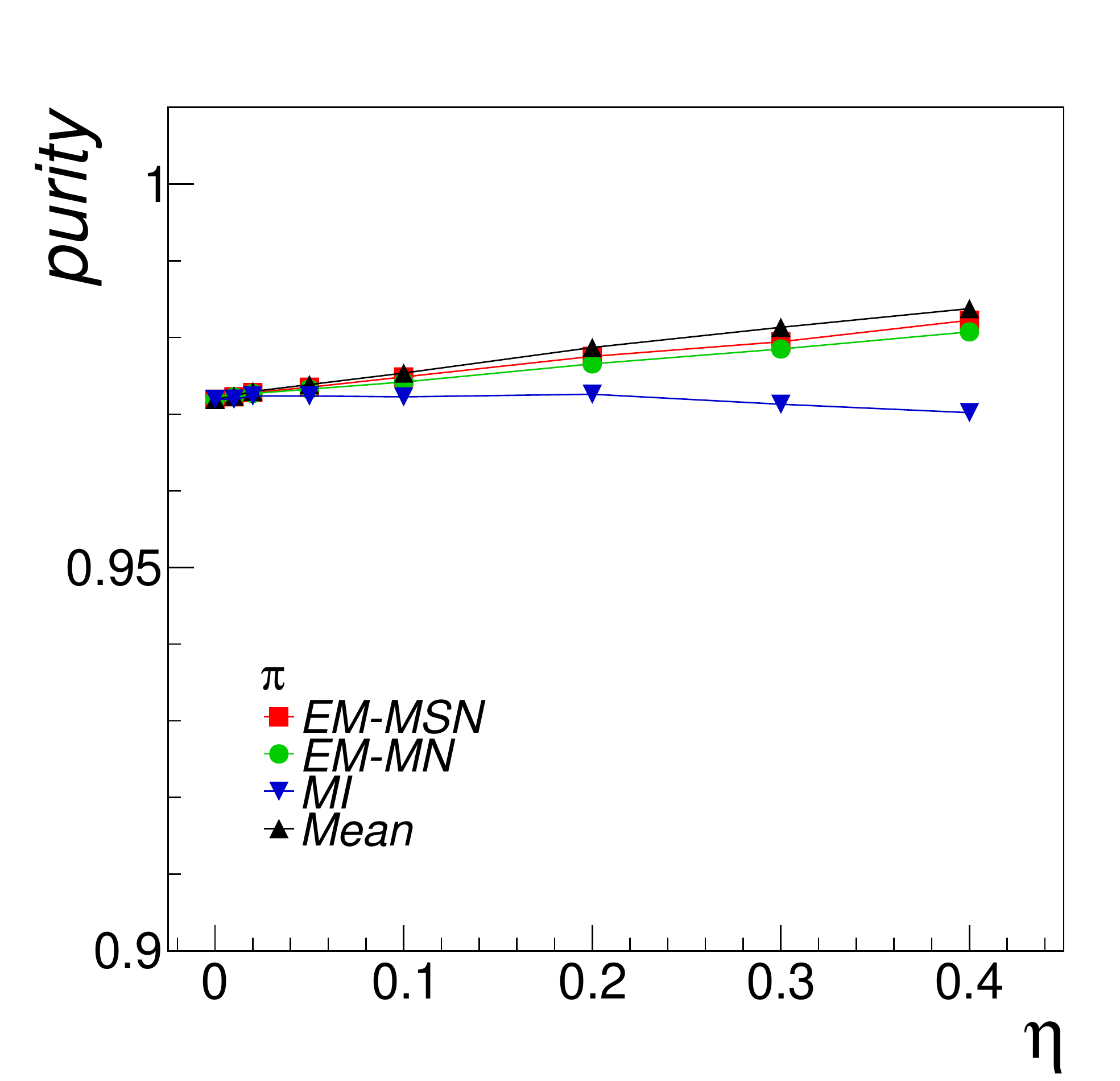}\label{PurityHighMomentumPlot1}}%
\hspace{-0.2cm}%
\subtable[$p$=(0.85-0.90) GeV/c - kaon]{\includegraphics[scale=0.28]{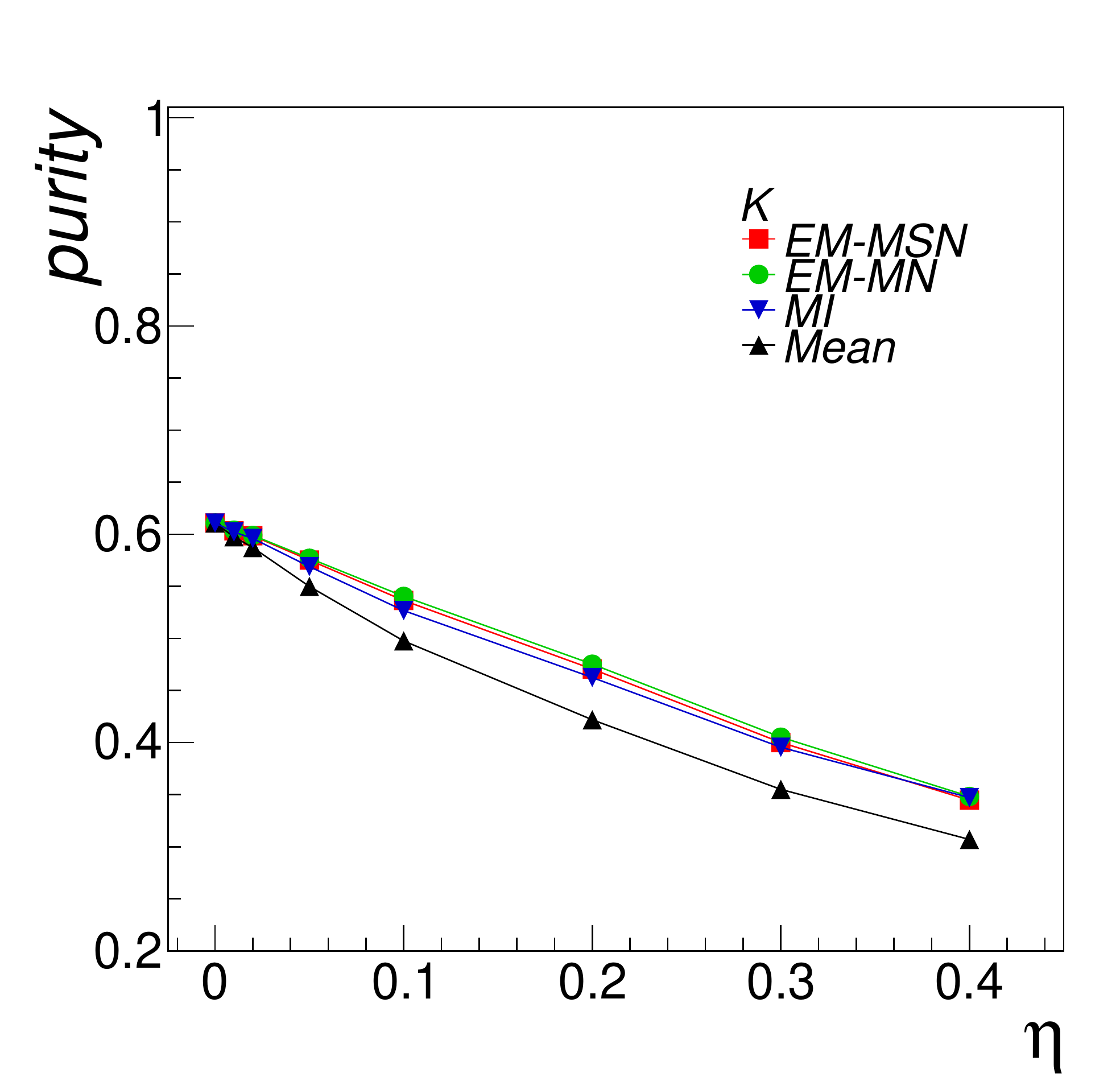}\label{PurityHighMomentumPlot2}}%
\hspace{-0.2cm}%
\subtable[$p$=(0.85-0.90) GeV/c - proton]{\includegraphics[scale=0.28]{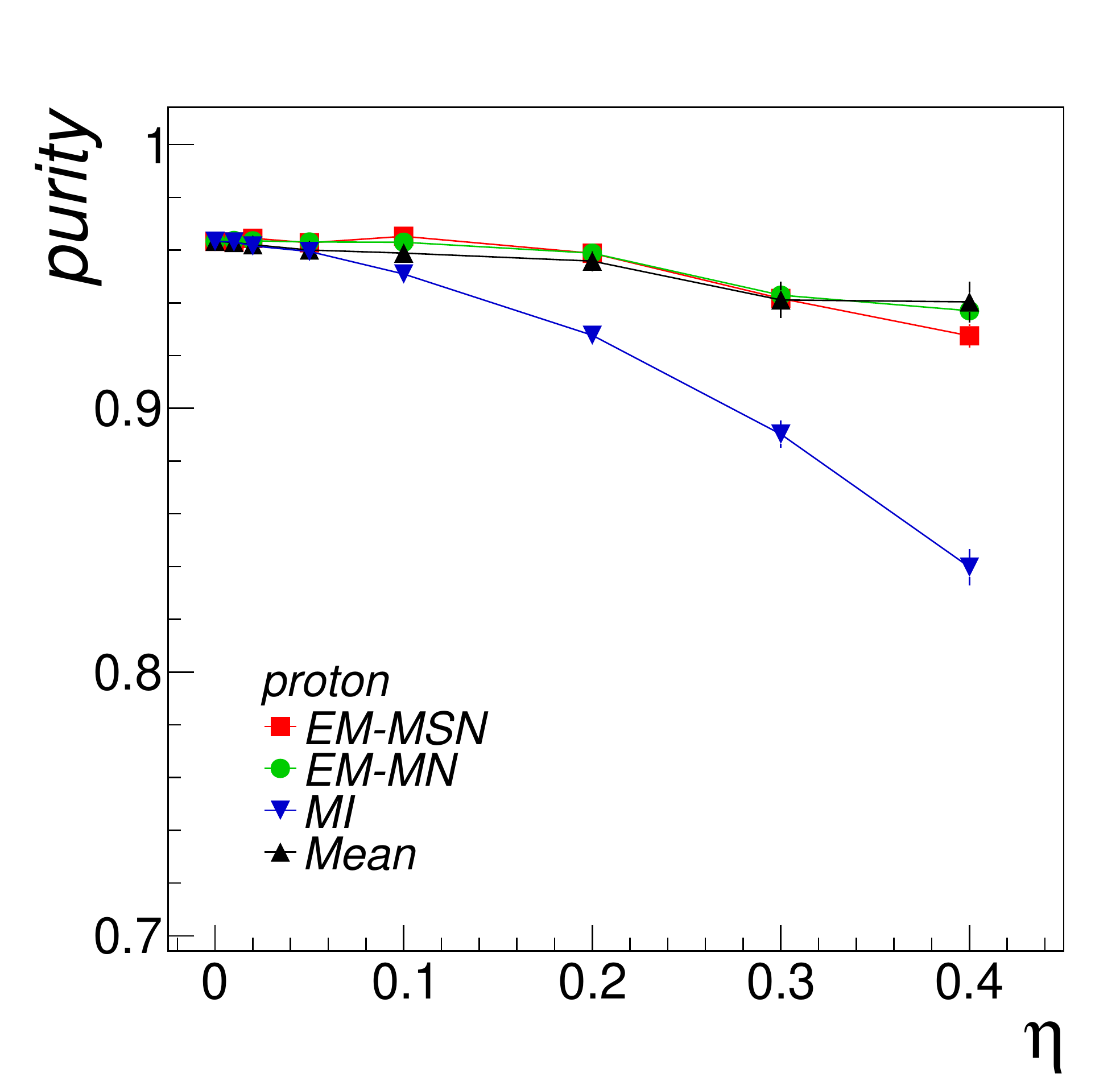}\label{PurityHighMomentumPlot3}}
\vspace{-0.2cm}
\caption{Classification purity for pions, kaons and protons, respectively shown from left to right obtained with different imputation methods
for three  momentum bins: (0.25-0.3) GeV/c (top panels), (0.55-0.60) GeV/c (middle panels) and  (0.85-0.90) GeV/c (bottom panels). 
Multiple imputation (MI) results
are shown with blue triangle markers, mean imputation with black markers, maximum likelihood with multivariate normal model (ML-MN) with green dots 
and maximum likelihood with multivariate skew-normal model (ML-MSN) with red squares.}
\label{PurityPlot}
\end{figure}

\section{Results}\label{ResultsSection}
In Fig. \ref{SampleImputationPlot} we report the scatter plot of the normalized  energy losses obtained in the first two detector layers 
for a sample test data set relative to a low particle momentum bin, namely (0.25-0.3 GeV/c). As specified before, instead of plotting the energy loss, 
the normalized (to that of kaons) logarithmic energy loss was reported. The full black dots indicate the complete (simulated) data, while the empty 
dots represent the imputed data obtained with the ML method for MN (Fig. \ref{SampleImputationPlot1}) and MSN (Fig. \ref{SampleImputationPlot2}) mixtures. 
The colored areas represent the contour levels of the MN and MSN bivariate model fitted to the data. Due to its asymmetry, the MSN model slightly 
better reproduces the data outside the region with the highest density with respect to the Gaussian model.

We performed the reconstruction analysis on the entire set of simulated data, assuming the energy losses from all detectors
as relevant observables
for the analysis. Fig. \ref{NNOutputPlot} shows the distributions of the neural network output for three different momentum bins, namely 
(0.25-0.30 GeV/c), where a good separation between the different species may be expected on the basis of their energy losses, 
and (0.55-0.60), (0.85-0.90) GeV/c, where a partial 
overlapping of the three species is expected, with a larger contamination (i.e. smaller purity) in each sample.
Proper cuts on such distributions have been applied and optimized for each momentum interval to classify the three species.

An overall comparison among the various imputation methods may be carried out observing the differences between true (simulated)  and
imputed data. Fig. \ref{Differences} shows the overall average quadratic differences (all the particle species and all six variables included)
as a function of the missing fraction of data and for three
particle momentum bins. The differences originating from the four
imputation methods are shown with different symbols. As it is clearly seen, at low momenta (where the three species are well separated) 
the largest differences are observed with the mean imputation method,
whereas the multiple imputation (MI) gives smaller values and the two ML methods give even smaller (and comparable) differences. 
The similarity between the normal and skew-normal ML imputation results is likely a consequence of the transformation of the original 
variables, which results in a more symmetric distribution with respect to the original Landau distribution.

In the results shown above, the mean is largely dominated by the most abundant species, in this case the pions, which explains the corresponding 
large differences observed in the small momentum bin, where the values of the energy losses for the three species at a given momentum are
sensibly different.
In the intermediate momentum bin, the
trend is still visible, even if the effect is reduced. Finally, at higher momenta, all the methods give comparable results in terms of 
differences between missing values and imputed values. 

The results reported in Fig.\ref{Differences} are an average over the various species.
However, the effect of the various imputation methods may be expected to be strongly different when a specific species is selected, 
due to their relative abundances.  When missing data are replaced by
data imputed by the various methods to build the neural network, this results in a different efficiency and purity of the different species.

In Fig. \ref{EfficiencyPlot} we report the average classification efficiencies for the three species (left panels: pions, 
middle panels: kaons, right panels: protons),
achieved by imputing the missing 
values according to the different imputation methods,
in three different momentum bins: (0.25-0.30) GeV/c (top panels), (0.55-0.60) GeV/c (middle panels) and (0.85-0.90) GeV/c (bottom panels). 
Multiple imputation results are shown in blue triangles, 
while simple mean imputation in black triangles. Maximum likelihood results are shown with red squares for the skew-normal model and with green dots for 
the normal model. 

As expected, the simple mean imputation provides a good identification only for the most abundant species, the pions, 
whereas for kaons and protons, the classification
capabilities are significantly deteriorated as the fraction of missing data increases, especially in the low momentum bin. 
The other imputation methods provides compatible results for low particle 
momenta, where the three species are perfectly separated. No significant loss of efficiency is observed even for a large fraction of missing data. 

At intermediate and higher momenta the various imputation algorithms provide comparable results in case of the most abundant species (pions), while the
multivariate approach, either in the normal model (ML-MN) or in the skew-normal model (ML-MSN) provide the best results - by a large factor -
in case of protons, which are the less abundant species.

In the high momentum bin, where an overlap occurs in the NN output  between the various species, the kaon efficiency may exhibit a 
different behaviour, with an increasing efficiency as a function of the missing data fraction, even with the mean imputation method.
This is explained as due to the increasing number of kaons passing the cut which discriminates kaons from pions. An increasing deterioration of the kaon
purity is also observed, as shown in Fig. \ref{PurityPlot}, which reports the corresponding results in terms of classification purity 
for the three species and the same momentum bins.

\section{Summary}\label{SummarySection}

We have quantitatively discussed different imputation methods to handle missing values in a data set where these values had to be used as input
neurons in an artificial neural network. Apart from the simple imputation method making use of the mean value, the Multiple Imputation
and the Maximum Likelihood Imputation, either as Normal or Skew-Normal models, were considered and tested against a simple test case, dealing
with the identification of different particle species through a multi-layer telescope detector in different momentum bins, 
corresponding to various degree of difficulty in the separation between the different species.

While it is easy to understand that missing values cannot be recovered by the simple mean imputation method, especially for the species which
are less abundant, since the imputed values are strongly influenced by the most abundant one, the detailed application of more sophisticated
imputation methods is not trivial, and requires different aspects to be taken into account. Methods based on the ML are generally believed to
provide more realistic results over the simple mean imputation or even the Multiple Imputation method. However, depending on the specific
application where the missing values need to be replaced by a realistic guess, different choices are possible, and in general a careful study
needs to be carried out before trusting the final results.

The test case considered in the present paper, more than providing a unique solution to this complex problem, was rather intended as a possibility
to discuss and implement the basic steps which  need to be addressed in most cases. The capability of any imputation method had to be checked
against simulated data first of all looking at the average differences between real and imputed data. However, dealing with the problem of
recognizing different species in an artificial neural network under various scenarios of increasing difficulty, it became clear that the
capabilities of the various methods are to be checked in each specific condition and in general the same method may behave slightly differently
when applied to different circumstances. Only after taking into account such aspects, may a specific imputation method be considered as a
reliable way to impute missing values in a data set. 

Concerning the application of the ML method to problems which exhibit intrinsically asymmetric distributions, as it is for example the Landau
distribution describing the specific energy loss of charged particles in thin layers, the ML with a skew-normal model should be derived for the
particular case of interest. This was done in this paper, and the detailed results are shown as an example in \ref{SkewNormalAlgorithmDerivation}. 
However, the possibility to proceed to a suitable transformation of variables in order to make the distributions of interest more similar to a normal
distribution, has also to be considered. It was shown in this paper that in such a case, the performance of the ML normal and ML skew normal
should be expected to be similar, as it was observed in our case.

Many of the considerations discussed in the present paper may also be applied to other physical problems of interest in which 
event statistics is a major concern. One of this is the mass identification of high energy primary cosmic rays in a hybrid detector, 
where missing values in a data set may arise from MCAR (i.e. inactive/unavailable areas for observables measured with surface stations), 
MAR (i.e. selection cuts on atmospheric parameters for observables recorded with fluorescence detectors) 
and MNAR (e.g. quality and fiducial cuts on same composition observables or on different shower variables directly correlated with them) mechanisms.

\section{Acknowledgements}\label{AcknowledgementsSection}

We warmly thank Prof. S.Ingrassia for useful discussions and previous collaboration works on the use of statistical methods applied to
real data sets.

\appendix%

\section{Learning Skew-Normal Mixture Models with missing data}\label{SkewNormalAlgorithmDerivation} 
Consider a sample of $N$ data $\bx_{i}$= ($\bx_1$,\dots,$\bx_{N}$) in which for a given observation $i$ we may have ignorable missing patterns, e.g. the 
data are missing at random (MAR) or completely at random (NMAR). We partition the data vector $\bx_i$ into the observed $\bx_{i,o}$ and the missing subvector 
$\bx_{i,m}$. Correspondingly, for the $i$-th observation the mixture parameters can be divided into an observed-missing part: ($\bxi_{ik,o}$, $\bxi_{ik,m}$),
($\bSigma_{ik,oo}$,$\bSigma_{ik,mm}$,$\bSigma_{ik,om}$), ($\bdelta_{ik,o}$,$\bdelta_{ik,m}$).\\
In the \emph{Expectation-Maximization} (EM) framework the $N$ observed data are considered incomplete and a set of variable indicators $\bZ=(\bz_1,\ldots,\bz_N)$
is introduced, such that $\bz_i=(z_{1i},\ldots,z_{Ki})$ (i=1,\dots N) with $z_{ki}$=1 if $\bx_i$ comes from the $k$-th component and $z_{ki}$=0 otherwise.
Following \cite{Zoubin1994}, missing features can be handled in the EM framework by regarding $\bx_{i,m}$ as additional missing variables as the indicators $z_{i}$. 
Using the stochastic representation of the skew-normal distribution, the complete log-likelihood can be written as:
\begin{align*}
\cL_c (\btheta)=& \cL_{\pi}(\pi_k) + \cL_{\xi\Sigma\delta}(\bxi_{k},\bSigma_{k},\bdelta_{k})\\
\mathcal{L}_{\pi}=& \sum_{i=1}^{N}\sum_{k=1}^{K}z_{ik}\log\pi_{k}\\%
\mathcal{L}_{\xi\Sigma\delta}=& \sum_{i=1}^{N}\sum_{k=1}^{K}\frac{z_{ik}}{2}\biggl[\log|\bSigma_{k}^{-1}|-(\bx_{i,o}-\bxi_{k,o}-\bdelta_{k,o}u_{i})^{T}
\bSigma_{k,oo}^{-1}(\bx_{i,o}-\bxi_{k,o}-\bdelta_{k,o}u_{i})+u_{i}^{2}\biggl]+\\%
+&\frac{z_{ik}}{2}\biggl[-(\bx_{i,m}-\bxi_{k,m}-\bdelta_{k,m}u_{i})^{T}
\bSigma_{k,mm}^{-1}(\bx_{i,m}-\bxi_{k,m}-\bdelta_{k,m}u_{i})\biggl]+\\%
+&\frac{z_{ik}}{2}\biggl[-(\bx_{i,o}-\bxi_{k,o}-\bdelta_{k,o}u_{i})^{T}
\bSigma_{k,mo}^{-1}(\bx_{i,m}-\bxi_{k,m}-\bdelta_{k,m}u_{i})\biggl]%
\end{align*}
where $\bU=(u_1, \ldots, u_N)$ are also unobservable latent variables.\\
In the $E$ step the EM algorithm requires to compute the expected value $Q(\bTheta|\bTheta^{(t)})= 
\mathbb{E}[\cL_c (\bTheta|\bX_{o},\bX_{m},Z,U)|\bX_{o},\bTheta^{(t)}]$ of the complete log-likelihood, 
which is then maximized in the $M$ step with respect to $\bTheta$ to obtain a new parameter estimate $\bTheta^{(t+1)}$.
This implies computing the following expectations:
\begin{align}
&\mathbb{E}(z_{ik}|\bx_{i,o})\equiv\tau_{ik}=\frac{\pi_{k}^{(t)}f(\bx_{i,o}|\btheta_{k,o}^{(t)})}{\sum_{k=1}^{K}\pi_{k}^{(t)}f_{k}(\bx_{i,o}|\btheta_{k,o}^{(t)})}\\
&\mathbb{E}(u_{i}|\bx_{i,o},z_{ik}=1)\equiv e_{1,ik}\\%
&\mathbb{E}(u_{i}^{2}|\bx_{i,o},z_{ik}=1)\equiv e_{2,ik}\\%
&\mathbb{E}(\bx_{i,m}|\bx_{i,o},z_{ik}=1)\equiv\hat{\bx}_{ik,m}\\%
&\mathbb{E}(\bx_{i,m}\bx_{i,m}^T|\bx_{i,o},z_{ik}=1)\\%
&\mathbb{E}(\bx_{i,m}u_i|\bx_{i,o},z_{ik}=1)
\end{align}

To compute the expectations relative to $u_i$ we need to consider the conditional distribution $u_i|(\bx_{i,o},z_{ik}=1)$. Following \cite{Lin2009}, it can be shown, 
by using the skew-normal hierarchical representation and the Bayes' Theorem, that $u_i|(\bx_{i,o},z_{ik}=1)$ follows a univariate truncated 
normal $\sim\mathcal{TN}(\mu_{ik,o},\sigma_{ik,o}^{2},0,+\infty)$:\\ 
\begin{align*}
&u_i|(\bx_{i,o},z_{ik}=1)\sim\mathcal{TN}(\mu_{ik,o},\sigma_{ik,o}^{2},0,+\infty)\\%
&\mu_{ik,o}=\bdelta_{ik,o}^T\bOmega_{k,oo}^{-1}(\bx_{i,o}-\bxi_{ik,o})\\
&\sigma_{ik,o}^{2}=1-\bdelta_{ik,o}^{T}\bOmega_{ik,oo}^{-1}\bdelta_{ik,o}
\end{align*}
The needed expectations are therefore given by the first and second moments of the truncated
normal distribution:
\begin{align*}
&e_{1,ik}\equiv\mathbb{E}(u_{i}|\bx_{i,o},z_{ik}=1)=\mu_{ik,o}+\sigma_{ik,o}\frac{\phi(\mu_{ik,o}/\sigma_{ik,o})}{\Phi(\mu_{ik,o}/\sigma_{ik,o})}\\%
&e_{2,ik}\equiv\mathbb{E}(u_{i}^{2}|\bx_{i,o},z_{ik}=1)=\mu_{ik,o}^{2}+\sigma_{ik,o}^{2}+\mu_{ik,o}\sigma_{ik,o}\frac{\phi(\mu_{ik,o}/\sigma_{ik,o})}
{\Phi(\mu_{ik,o}/\sigma_{ik,o})}%
\end{align*}

To compute the additional expectations involving the observed $\bx_{ik,o}$ and missing data $\bx_{i,m}$
we recall from \cite{Lin2010} the following results concerning the involved conditional distributions:
\begin{align*}
&\bx_{i,o}|(u_i,z_{ik}=1)\sim\mathcal{N}_{p_{i,o}}(\bxi_{ik,o}+\bdelta_{ik,o}|u_i|,\bSigma_{ik,oo})\\
&\bx_{i,m}|(\bx_{i,o},u_i,z_{ik}=1)\sim\mathcal{N}_{p_{i,m}}(\bxi_{ik,m}+\bdelta_{ik,m}|u_i|+\bSigma_{ik,mo}\bSigma_{ik,oo}^{-1}(\bx_{i,o}-\bxi_{ik,o}-
\bdelta_{ik,o}|u_i|),\bSigma_{ik,mm}-\bSigma_{ik,mo}\bSigma_{ik,oo}^{-1}\bSigma_{ik,om})\\ 
&u_i|(\bx_{i,o},z_{ik}=1)\sim\mathcal{TN}(\bmu_{ik,o},\sigma_{ik,o}^2,0,+\infty)
\end{align*}
On the basis of such results, by employing the law of iterative expectations and of total covariance, it can be demonstrated that the expectations to be computed are 
given by:
\begin{align*}
&\hat{\bx}_{ik,m}\equiv\mathbb{E}(\bx_{i,m}|\bx_{i,o},z_{ik}=1)=\bxi_{ik,m}+\bdelta_{ik,m}e_{1,ik}+\bSigma_{ik,mo}\bSigma_{ik,oo}^{-1}(\bx_{i,o}-\bxi_{ik,o}-
\bdelta_{ik,o}e_{1,ik})\\%
&\mathbb{E}(\bx_{i,m}\bx_{i,m}^T|\bx_{i,o},z_{ik}=1)=\hat{\mathbf{C}}_{ik,mm}+
\hat{\bx}_{ik}\hat{\bx}_{ik}^T\\%
&\mathbb{E}(\bx_{i,m}u_i|\bx_{i,o},z_{ik}=1)= e_{1,ik}\tilde{\bx}_{ik,m}
\end{align*}
where $\hat{\mathbf{C}}_{ik,mm}=\bSigma_{ik,mm}-\bSigma_{ik,mo}\bSigma_{ik,oo}^{-1}\bSigma_{ik,om} + 
(e_{2,ik}-e_{1,ik}^2)(\bdelta_{ik,m}-\bSigma_{ik,mo}\bSigma_{ik,oo}^{-1}\bdelta_{ik,o})(\bdelta_{ik,m}-\bSigma_{ik,mo}\bSigma_{ik,oo}^{-1}\bdelta_{ik,o})^{T}$ 
denotes the covariance of the missing values $\bx_{i,m}$ and $\tilde{\bx}_{ik,m}=\bxi_{ik,m}+\bdelta_{ik,m}e_{2,ik}/e_{1,ik}+
\bSigma_{ik,mo}\bSigma_{ik,oo}^{-1}(\bx_{i,o}-\bxi_{ik,o}-\bdelta_{ik,o}e_{2,ik}/e_{1,ik})$.\\
Using the above results we computed the expected value of the complete likelihood and applied the $M$ step, e.g. setting the derivatives with respect to the 
mixture parameters $\bTheta$ equal to zero and solving for $\bTheta$. After a lengthy algebra calculation we determined the following parameter update rules:
\begin{align*}
&\tau_{ik}^{(t+1)}= \frac{\pi_{k}^{(t)}f(\bx_{i,o},\btheta_{k,o}^{(t)})}{\sum_{k=1}^{K}\pi_{k}^{(t)}f_{k}(\bx_{i,o},\btheta_{k,o}^{(t)})}\\%
&\pi_{k}^{(t+1)}= \frac{1}{N}\sum_{i=1}^{N}\tau_{ik}^{(t+1)}\\%
&\bxi_{k}^{(t+1)}= \frac{\sum_{i=1}^{N}\tau_{ik}^{(t+1)}(\hat{\bx}_{ik}-\bdelta_{k}^{(t)}e_{1,ik}^{(t)})}{\sum_{i=1}^{N}\tau_{ik}^{(t+1)}}\\%
&\bdelta_{k}^{(t+1)}= \frac{\sum_{i=1}^{N}\tau_{ik}^{(t+1)}e_{1,ik}^{(t)}(\tilde{\bx}_{ik}-\bxi_{k}^{(t+1)})}{\sum_{i=1}^{N}\tau_{ik}^{(t+1)}e_{2,ik}^{(t)}}\\%
&\bSigma_{k}^{(t+1)}= \frac{\sum_{i=1}^{N}\tau_{ik}^{(t+1)}  (\hat{\bx}_{ik}-\bxi_{k}^{(t+1)}) (\hat{\bx}_{ik}-\bxi_{k}^{(t+1)})^{T} } 
{\sum_{i=1}^{N}\tau_{ik}^{(t+1)}} - 
\frac{\sum_{i=1}^{N}\tau_{ik}^{(t+1)} e_{1,ik}^{(t)}[\bdelta_{k}^{(t)}(\tilde{\bx}_{ik}-\bxi_{k}^{(t+1)})^{T}+(\tilde{\bx}_{ik}-\bxi_{k}^{(t+1)})
\bdelta_{k}^{T (t+1)}]}
{\sum_{i=1}^{N}\tau_{ik}^{(t+1)}}+\\%
+& \frac{\sum_{i=1}^{N}\tau_{ik}^{(t+1)} e_{2,ik}\bdelta_{k}^{(t+1)}\bdelta_{k}^{T (t+1)}]}
{\sum_{i=1}^{N}\tau_{ik}^{(t+1)}}+
\frac{\sum_{i=1}^{N}\tau_{ik}^{(t+1)}\hat{\mathbf{C}}_{ik}}{\sum_{i=1}^{N}\tau_{ik}^{(t+1)}}
\end{align*}

\end{document}